\documentclass[%
 aip,
 amsmath,amssymb,
 reprint,%
]{revtex4-1}

\usepackage{graphicx}
\usepackage{dcolumn}
\usepackage{bm}

\usepackage[utf8]{inputenc}
\usepackage[T1]{fontenc}
\usepackage{mathptmx}
\usepackage{etoolbox}
\usepackage{natbib}
\setcitestyle{numbers}
\setcitestyle{square}

\makeatletter
\def\@email#1#2{%
 \endgroup
 \patchcmd{\titleblock@produce}
  {\frontmatter@RRAPformat}
  {\frontmatter@RRAPformat{\produce@RRAP{*#1\href{mailto:#2}{#2}}}\frontmatter@RRAPformat}
  {}{}
}%
\makeatother
\begin{document}

\preprint{AIP/123-QED}

\title[]{Polyelectrolyte Complexation of Two Oppositely Charged Symmetric Polymers: A Minimal Theory}
\author{Soumik Mitra}
 \affiliation{$^1$ Department of Physical Sciences, Indian Institute of Science Education and Research Kolkata, Mohanpur 741246, India}
\author{Arindam Kundagrami}%
%
 \altaffiliation[Also at ]{Centre for Advanced Functional Materials, Indian Institute of Science Education and Research Kolkata, Mohanpur 741246, India}
 \affiliation{$^1$ Department of Physical Sciences, Indian Institute of Science Education and Research Kolkata, Mohanpur 741246, India
}

\begin{abstract}
Interplay of Coulomb interaction energy, free ion entropy, and conformational elasticity is a fascinating aspect in polyelectrolytes (PEs). We develop a theory for complexation of two oppositely charged PEs, a process known to be the precursor to the formation of complex coacervates in PE solutions, to explore the underlying thermodynamics of complex formation, at low salts. Explicit calculation of the free energy of complexation and its components indicates that the entropy of free counterions and salt ions and the Coulomb enthalpy of bound ion-pairs dictate the equilibrium of PE complexation. This helps decouple the self-consistent dependency of charge and size of the uncomplexed parts of the polyions, derive an analytical expression for charge, and evaluate the free energy components as functions of chain overlap. Complexation is observed to be driven by enthalpy gain at low Coulomb strengths, driven by entropy gain of released counterions but opposed by enthalpy loss due to reduction of ion-pairs at moderate Coulomb strengths, and finally prohibited by enthalpy loss at higher Coulomb strengths. Phase diagrams are constructed which identify the stability of fully-, partially- and un-complexed states as functions of electrostatic strength. Thermodynamic predictions from our model are in good quantitative agreement with simulations in literature, and may motivate simulations and experiments at higher electrostatic strengths at which complexation is found to be unfavourable.
\end{abstract}

\maketitle

\section{Introduction}

Polyelectrolyte complexes (PECs) formed by the association of  oppositely charged 
polylectrolyes (PE) are a class of extremely popular materials widely investigated by polymer scientists and engineers. The volume of experimental\cite{kabanov1985,dautzenberg2002,spruijt2010macro,shamoun2012,perry2014,salehi2015,lutkenhaus2015,fu2016,
schlenoff2017,ali2018,tirrell2018,tirrell2020,tirrell2021}, simulational\cite{winkler2002,hayashi2004,zhang2005,zhaoyang2006,fredrickson2007,larson2009,semenov2012,lytle2016,chakrabarti2018,
whitmer2018,whitmer2018macro,rathee2018,shakya2020,yethiraj2020}, and theoretical\cite{voorn1957,deVries2006,oskolkov2007,perry2015,pablo2016,salehi2016,lytle2017,potemkin2017,
rumyantsev2018,zhen-gang2018,adhikari2018,ong2019,rumyantsev2019} research done on PECs is overwhelming and exciting. In the polyelectrolyte coacervates, which form due to complexation of multiple pairs of PE chains, the phase separation entails both types of polyions ending up in the same phase, while the coexisting phase contains a low concentration of polymers (supernatant). Generally, PEs dissociate in aqueous solution to release counterions of opposite signature,  
PEC formation results in the release of almost all counterions from the PE chain backbones. Simultaneously, during the complexation process, there is a redistribution of bound ion-pairs, 
starting from the counterion-monomer pairs leading to pairs of oppositely charged monomers. The flexibility of the polymer chains, on the other hand, offers additional entropy to the system. In addition, the presence of small ions in the  solution screens all electrostatic effects at various degrees. This complex interplay of entropy and enthalpy in the presence of Coulomb screening allows PECs to play a significant role in nature, making it a part of many biological phenomena, such as protein-polyelectrolyte association\cite{DeRouchey2005,daSilva2009,kayitmazer2013}. PECs are extremely promising for biomedical applications including drug transfer, gene transfection, and gene therapy. Hosting the interplay of energy, entropy, and elasticity at room temperatures, PECs are fascinatingly rich candidate structures to verify fundamental principles of soft matter and polymer physics. 

In early days PE complexation was primarily interpreted in terms of electrostatic attraction between oppositely charged polyions\cite{kudlay2004,spruijt2010macro}. The Voorn-Overbeek theory (VOT)\cite{voorn1957}, that ignores charge connectivity, combined the Flory-Huggins mixing entropy and Debye-H\"uckel\cite{mcquarrie2000} energy of ionic solutions to explain the coexisting phase behaviour in the solution of oppositely charged polymers. Despite distinct limitations VOT remains one of the favourites, if not the most, of the community, whereas the theory has been extended later by including chain connectivity, counterion effects, or excluded volume interactions or it is referred by experimentalists while they put forward simple model calculations to fit their experimental data. Several modifications of VOT came up for the next 60 years\cite{veis1960,biesheuvel2004,oskolkov2007,perry2015,pablo2016,salehi2016}, and so did several other theories with varied levels of sophistication\cite{joanny2001,delacruz2004,kudlay2004,fredrickson2007,fredrickson2012,perry2015,lytle2017,
fredrickson2017}. 

From the earliest analysis of Voorn and Overbeek\cite{voorn1957} and early models\cite{veis1960,nordmeier1999} of complex formation/coacervation, up until the recent advanced approaches\cite{joanny2001,zhang2005,spruijt2010macro,perry2015,pablo2016,potemkin2017,lytle2017}, the emphasis was on electrostatic interactions and its correlations at various scales in the multi-component coacervates, but not always with specific analysis of the roles of enthalpy and entropy of different 
charged species. Until around ten years ago, only a handful of literature\cite{michaels1965,djadoun1977,tsuchida1980,tianaka1980,dautzenberg2002,zhaoyang2006} considered explicit
entropy of released counterions. Several approaches, however, 
in recent times have explored\cite{beltran2012,perry2015,salehi2016,fu2016,muthu2017,lytle2017,adhikari2018,whitmer2018,
whitmer2018macro} the comparative roles of entropy of counterion release and 
electrostatics of multi-component charged species in providing the thermodynamic drive for coacervate formation and phase separation for solutions of PECs.  Reviews  \cite{deVries2006,veis2011,gucht2011,srivastava2016,sing2017,muthu2017,meka2017,sing2020,rumyantsev2021} highlighted the question, and progressively shed better light recently. Among the theoretical approaches beyond mean field models, Random Phase Approximation (RPA)\cite{borue1988,borue1990,delacruz2003,delacruz2004,kudlay2004,mahdi2000,joanny2001}, Transfer Matrix methods\cite{lytle2017}, molecular simulations \cite{winkler2002,hayashi2004,larson2009,lytle2016,rathee2018,whitmer2018,whitmer2018macro,chen2022} , as well as field theoretic simulations\cite{fredrickson2007,fredrickson2012,fredrickson2017} (FTS) studied several other aspects of this fascinatingly rich process, but a strong emphasis shifting towards the counterions
and related enthalpy-entropy interplay only came recently\cite{fu2016,rathee2018,whitmer2018,whitmer2018macro,yethiraj2020}. Sing and Perry\cite{perry2015} incorporated strong charge correlations arising from counterion condensation by applying the Ornstein-Zernike formalism in the PRISM model. Effects of charge distribution on the PE backbone\cite{dautzenberg2002,oskolkov2007,whitmer2018macro} and profound effect of 
added salt and the salt partitioning behavior in the coexisting phases for symmetric mixtures of polyions\cite{dautzenberg2002,
kudlay2004,tirrell2012,perry2014,zhengang2018,dePablo2018,ali2018,tirrell2018,tirrell2020,shakya2020} are a few other key aspects investigated. It is now well established that RPA has limitations in dealing with strongly charged systems, since it does not take into account of the energy of bound ion-pairs in its entirety. In FT, utmost care and consideration are given to accurately calculate the charge correlations in the bulk solution of oppositely charged polyions, but it does not quantitatively accomodate for the counterions. The common feature of all theoretical approaches
has been that they considered bulk solutions of polyion pairs, 
and hence had to deal with both intra- and inter-pair interactions and charge correlations that are signatures of a complex many-body problem, leading to inherently complex integral equation theories or FT methods. 
 
Experimental investigations \cite{record1978,kabanov1985,dautzenberg2002,spruijt2010macro,perry2014,salehi2015,lutkenhaus2015,schlenoff2017,
ali2018,tirrell2018,dePablo2018,rumyantsev2018,ali2019,wang2019,pde2020,tirrell2020} on PECs date back to more than half a century. Understandably, solutions of PECs and phase coexistence of coacervates and supernatants therein have been the preferred systems of study, compared to only a few pairs of chains in high dilutions. One may note that the role of free ion entropy is most prominently manifested in high dilutions, but the regime has experimental challenges\cite{pde2020}.
 Isolated early efforts\cite{michaels1965,djadoun1977,tsuchida1980,tianaka1980} stressed on the effect of dissociated counterions in the formation of PECs. Very recently several experiments \cite{fu2016,meka2017,wang2019,tirrell2018} have stressed on the counterion entropy as the primary thermodynamic drive for PEC formation. Tirrell and co-workers explored rheological properties and solvent effects\cite{tirrell2012,tirrell2013,tirrell2014,tirrell2018}, and identified the role of counterion release and entropy of free ions in several 
observations\cite{tirrell2012,tirrell2018,tirrell2020,tirrell2021}. Schlenoff and co-workers\cite{fu2016,schlenoff2017,wang2019}
emphasized on the role of entropy as the driving force and of electrostatic interactions as a prohibitor in PEC formation. Similar stress on the relevance of counterion entropy is found in simulations\cite{zhaoyang2006,larson2009,semenov2012,rathee2018,whitmer2018macro}. 

Although significant advances have been made in the last decade to settle the pressing issue of the thermodynamic driving 
(or opposing) forces of complexation, an understanding based on simple models is still rare, in our opinion. The parameter space for such a complex, multi-component process is vast, making such driving forces dependent on density of PEs, temperature, dielectric constant of the solvent, solvent quality, ionic strength of the solution, flexibility of the chains, charge densities etc. The Langevin simulation of two oppositely charged PEs
with explicit counterions by Ou and Muthukumar\cite{zhaoyang2006} unambiguously showed that entropy of counterion release is favourable to complexation, whereas electrostatic
enthalpy, mainly arising due to bound ion-pairs, may or may not be, depending on the electrostatic interaction strength as compared to thermal energy (also observed
by Rathee et al.\cite{rathee2018,whitmer2018macro} later). Dzubiella \textit{et. al.}\cite{dzubiella2016} simulated the potential of mean force between two oppositely charged symmetric PEs sliding onto each other, and calculated explicit counterion entropy and enthalpy of complexation. It is amply observed\cite{semenov2012,priftis2012,vitorazi2014,chang2017,yethiraj2020,chen2022} that the formation of a coacervate comprises of two significant steps - 1) the ion pairing between oppositely charged polyions to form the so called 'soluble' complexes and 2) the subsequent separation of such pairs into rich and poor polymer phases, namely the coacervate and supernatant, respectively. In addition, traditional studies reveal that free ion entropy (ideal gas or mixing type) and bound ion-pair energy dictate the equilibrium structure and distribution of ions in several types of PE systems (single PE chains in dilute solution\cite{joanny1990,winkler1998,muthu2004,arindam2010,mitra2017}, PE solutions\cite{joanny1996,kramarenko2002,hua-arindam2010}, PE gels\cite{khokhlov1994,hua-mitra2012,swati2015,swati2017} etc.). It should be worthwhile to investigate this particular aspect of charged polymers in the analysis of equilibrium properties of PECs, 
especially with a very simple model.
 
It is quite evident that the complex formation between two oppositely charged polyions is an important primary step in the study of coacervation, exploring which based on the energetics of the process is of interest. Therefore, a basic analytical theory of complexation of two PE chains in a dilute salty solution, taking into account of all entropic and enthalpic contributions from the 
small ions and polyions, is still in order. 
A few questions that need answers from the theoretical model are: If the complexation proceeds through a gradual overlap of oppositely charged monomers, would the thermodynamics (the free energy) be monotonically favourable, or one shall have partial complexation depending on certain electrostatic interactions (set by temperature, dilectricity of the solvent, ionic strength etc.)? At which conditions is the complexation fully disfavoured and the chains prefer to stay separated?  What is the comparative role of enthalpy gain due to bound ion-pairs and entropy gain due to released counterions, how does it change with ambient conditions, and how does it compare with other contributions? What role is played by the chain elasticity before and after complexation?  

To answer the above questions, we have proposed a minimal theory for the complexation of two symmetric,
fully-ionizable (strong), and oppositely charged PEs
in a dilute solution with low salt.
The primary aim is to explore and quantify the comparative roles of the ideal gas entropy of 
free ions and electrostatic energy of bound ion-pairs in regulating the equilibrium structure and charge, and identify thermodynamic forces and factors which promote or prohibit the complexation.  The two PE chains are assumed to form a complex with one-to-one monomer 
mapping (the ladder model\cite{michaels1965,winkler2002,hayashi2004,zhaoyang2006,semenov2012,dzubiella2016,chen2022}). The free energy is constructed following the single chain theory of Edwards path integral approach within the uniform spherical expansion constraint\cite{edwards1979,muthu2004,arindam2010}, and its contributions come from the entropy of free ions and conformations of chains and electrostatic interaction among charged species. Explicit calculation of the free energy components and their loss or gain after complexation, along with suitable approximations leading to analytical expressions, help assess the thermodynamic driving and opposing forces of PE complexation of simple homopolymers. The basic findings of our theory, especially for the enthalpy and entropy of complexation, number of released counterions, and the potential of mean force (free energy in our theory) at various electrostatic interaction strengths, are compared quantitatively with simulations available in the literature. A phase diagram as function of electrostatic strength is constructed, which indicates an energetically disfavoured complexed state at high strengths.

\section{Model}

\begin{figure}[h]
\centering
\includegraphics[height=9.0cm,width=8.0cm]{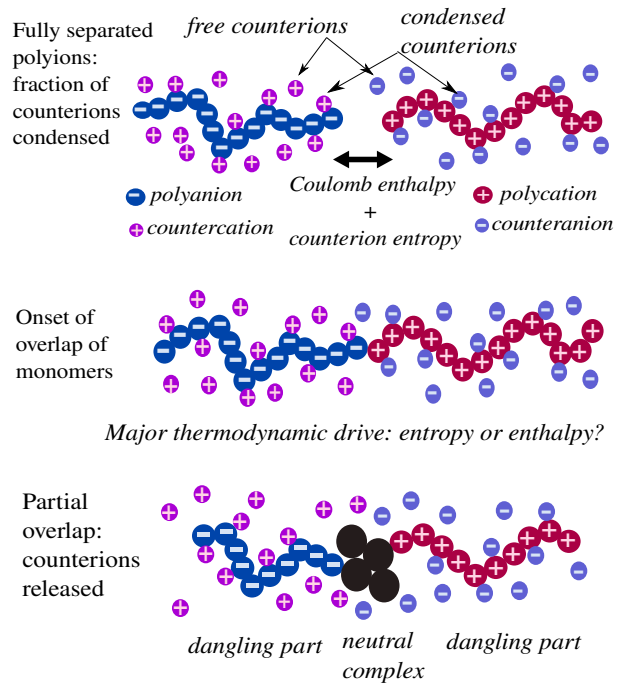}~~~
  \caption{Schematic of two sliding, symmetric PE chains overlapping to form a complex: The oppositely charged chains slide along each other as the monomers form bound ion-pairs, releasing 
  their counterions. In the dangling chain parts, the monomers remain partially compensated, as in a single isolated PE chain.}
  \label{overlap}
\end{figure}

\begin{figure}[h]
\centering
\includegraphics[height=5.8cm,width=8.0cm]{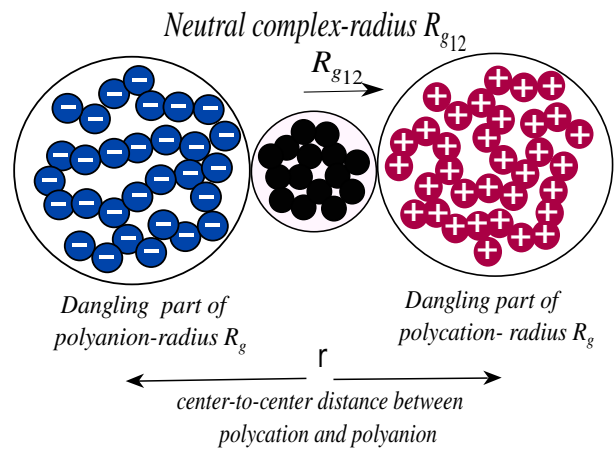}~~~
  \caption{Depiction of the two overlapping PE chains within the spherical model: The
  polyanion sphere (blue) and the polycation sphere (red) are intervened by the neutral complex 
  (black) made of monomer pairs.}
  \label{interactions}
\end{figure}

$~~~$Two oppositely charged, fully ionizable, linear flexible polyelectrolyte (PE) chains in high dilution are modeled as self-avoiding walks, and with electrostatic interactions (Fig. \ref{overlap}). The polycation (PC) and polyanion (PA) chains are assumed to have the same length and same number of monomers of the same 
size (the "symmetric" case), and they are assumed to complex commensurately, following the ladder model, or one-to-one monomer-mapping\cite{zhaoyang2006,dzubiella2016}. In an intermediate state of complexation, an equal number of monomers from each polyion form bound ion-pairs and get "complexed", and the intermediate complex has three parts - two {\it dangling}, uncomplexed parts of PA and PC on
either side, and a {\it complexed} collection of pairs of monomers coming from PA and PC in between them.  
The charged monomers of each dangling chain have oppositely charged counterions some of which are condensed on to the chain backbone, and the rest are free in the solution. All counterions of the complexed monomers, for the intermediate complex, are free.  Consequently, there remains mutual electrostatic interaction between the dangling parts of the two chains, as well as between monomers in the same chain, which is screened by the ionic solution made of free counterions and salt ions. The PE chains are assumed to undergo conformational changes, i.e., to expand or shrink, in response to their individual charge contents within the uniform expansion model approximation\cite{edwards1979,muthu2004,arindam2010}, in which each chain is  encapsulated in a hypothetical sphere of radius equal to the radius of gyration of the chain (Fig. \ref{interactions}). The pseudo-neutral complex made of bound-pair monomers is also taken to be a sphere, but with no electrostatic interactions. 

Starting from two completely separated PA and PC each with $N$ monomers, the process of complexation commences with formation of a single pair of monomers from either chain, forming one neutral `monomer' for the complex. For a general time instant, equal number, $n$, of monomers from both PE chains form a neutral complex of $n$ `monomers', leaving the dangling parts of the chains with $(N-n)$ monomers. This overlap process continues till the overlap number $n$ equals $N$, at which point all the monomers of the two chains form bound monomer-pairs that constitute the complete neutral complex of $N$ `monomers'. 

Although the complexation is a kinetic process, we adopt the quasistatic assumption, such that the size and charge of both dangling ends and the distribution of free ions in solution are assumed to equilibrate at each step of overlap. Eventually, at each step, we obtain $R_g$ and $f$ through self-consistent double minimization of the model free energy. The work studies the variation of these two basic attributes of the dangling chains, 
along with the different free energy components, as functions of overlap ($\lambda=n/N$). Further, we calculate
the free energy difference (free energy of complexation) between the fully separated ($\lambda=0$) and fully complexed ($\lambda=1$) states of the two chains, that determines the thermodynamic driving force for complexation. By analyzing
the individual components of the free energy, we identify their contributions to the thermodynamic driving force and quantify their values for all degrees of overlap.  

\subsection{The Free Energy} 

The free energy of the dangling parts of both chains PA and PC is calculated within the uniform expansion model using Edwards Hamiltonian applicable for a single isolated polyelectrolyte in dilute solution\cite{edwards1979,muthu2004,arindam2010}, and is suitably modified to address
the more involved system of two PE chains and an intermediate neutral complex we have at hand. The Edwards Hamiltonian and related free energy have been used to describe the polymer characteristics and related thermodynamical variables in numerous problems of charged PE systems \cite{muthu2004,dua2005,dua2007,kundu2014}, solutions \cite{vilgis1991}, and to match experimental results\cite{beer1997,loh2008}. The free energy consists of the contributions from the (i) translational entropy of counterions condensed on monomers, (ii) translational (ideal gas) entropy of the released and free counterions and salt ions in the solution, (iii) fluctuations in density of all dissociated ions (Debye-H\"uckel), (iv) attractive Coulomb energy of ion-pairs formed due to both condensation of counterions on the chain backbones and oppositely charged monomers from PA and PC, and (v) conformational entropy of the flexible chains (dangling
PA and PC and the neutral complex). In addition, specially for this complexation scenario of two oppositely charged macroions, one needs to consider the contribution from the (vi) electrostatic interaction between the two dangling PE chains from PA and PC, respectively, as well.

$~~~$At an intermediate state of overlap of two polyions, the complex consists of three polymer chains, two of which are the dangling PC and PA, respectively, of length $N-n$, and one is the complexed and neutralized chain of length $n$ (Fig. \ref{interactions}). The overall volume of this extremely dilute solution (made of the polymers, ions, and solvent) is $\Omega$, whereas the number of lattice points may be chosen as $\Omega/l^3$, where $l$ is the dimension of the smallest unit of volume, typically chosen to be the size of a monomer. These charged macromolecules (PA and PC) ionize in aqueous medium to release monovalent counterions of the opposite signatures in the solution. Considering $M$ counterions condensed on $N-n$ ionizable monomers of the dangling parts of each chain, the degree of ionization of each dangling part will be $f=1-\frac{M}{N-n}$. If we assume that the local dielectric environment for the two chains are similar, with equal monomer sizes the prevailing symmetry results in equal degree of ionization for them, i.e., $f_1=f_2=f$. The size of a polymer is given by the radius of gyration $R_g$, which shall also be the same for the PA and PC, once we assume that the intrinsic excluded volume parameter, $w$, is the same for them. Further, let $c_s$ be the number density of molecules of an externally added salt that dissociates into $n_+$ cations and $n_-$ anions, where, $c_s=n_+/\Omega$ $=$ $n_-/\Omega$=$n_s/\Omega$. For simplicity, we assume the salt cations (and anions) to be of the same type of the countercations (and counteranions) of the PA and PC, respectively. Therefore, at a certain value of overlap $\lambda=n/N$, $N-M+n_s$ monovalent cations of the same type and equal number of anions of the same type 
remain free in the solution. Generally, like in any polyelectrolyte system, there exists a mutual dependency between the charge ($f$) and and size ($R_g$) of the chains\cite{muthu2004,arindam2010}, and the total free energy of the system needs to be self-consistently minimized in terms of these two variables to reach the equlilibrium. The overall monomer number density for the dangling parts can be written as $\rho=(N-n)/\Omega$. We create the dimensionless variables $\tilde{c}_s=c_sl^3$ and $\tilde{\rho}=\rho l^3$. 

There are $W={{N-n} \choose M}$=$(N-n)!/M!(N-n-M)!$ ways $M$ condensed counterions may distribute itself along the backbone of each chain. Hence, the entropy is $k_B \log W$, and the free energy contribution due to the translational entropy of condensed counterions is
\begin{equation}
\frac{F_1}{k_B T}=2(N-n)\left[f\log f + (1-f)\log(1-f)\right],
\label{free1}
\end{equation}
where $k_B, T$ are the Boltzmann constant and temperature, respectively.
The translational or free-volume entropy of the free counterions plus all other 
free ions (in this case the ones coming from the dissociated salt) in solution is $k_B \log [(\Omega/l^3)^{\sum n_i}/\Pi n_i !] = - k_B \Omega[\sum c_i \log (c_i l^3)-\sum c_i]$, where $n_i$ is the number of ions of species $i$ and $c_i=n_i/\Omega$. Considering $N-M+n_s$ free ions for each species, the free energy contribution from free-ion entropy turns out to be 
\begin{align}
\frac{F_2}{k_B T}= & 2 (N-n) \left\lbrace \left[f+\frac{n}{N-n}+\frac{\tilde{c}_s}{\tilde{\rho}}\right]\log\left[\tilde{\rho} f + \frac{\tilde{\rho} n}{N-n}+\tilde{c}_s\right]\right. \nonumber\\
&\left.  -\left[f+\frac{n}{N-n}+\frac{\tilde{c}_s}{\tilde{\rho}}\right] \right\rbrace.
\label{free2}
\end{align}

The free energy contribution arising due to the fluctuations in the densities of the free counterions and coions, including the salt ions (the same set of free ions applicable for
$F_2$), is given by the Debye-H\"uckel theory, in the form $-k_B T \Omega \kappa^3/12\pi$, where $\kappa$ is the inverse Debye screening length, given by $\kappa^2=4 \pi l_B \sum_i Z_i^2 n_i/\Omega$,
where $Z_i$ is the valency of ion species $i$, and $l_B=e^2/(4\pi \epsilon_0 \epsilon k_B T)$ is the Bjerrum length, where $e, \epsilon_0$, and $\epsilon$ are, respectively, the electronic charge, dielectric permittivity of vacuum, and the bulk dielectric constant of the solvent. Again, considering $N-M+n_s$ free ions for each monovalent species one gets 
\begin{equation}
\tilde{\kappa}^2=8 \pi \tilde{l}_B \left[\tilde{\rho} f + \frac{n}{N-n} \tilde{\rho} + \tilde{c}_s\right].
\label{kappa}
\end{equation}
Using this value of $\kappa$, we have the fluctuation energy of the form of
\begin{equation}
\frac{F_3}{k_B T}=-\frac{\Omega\kappa^3}{12\pi}=-\frac{2}{3}\sqrt{\pi} \tilde{l}_B^{3/2} \frac{(N-n)}{\tilde{\rho}} \left(2\left[\tilde{\rho} f + \frac{n}{N-n}\tilde{\rho} +\tilde{c}_s\right]\right)^{3/2},
\label{free3}
\end{equation}
where $\Omega/l^3$ has been replaced by $(N-n)/\tilde{\rho}$.

A condensed counterion on the polymer backbone forms an ion-pair with its nearest charged monomer. Within the complex the oppositely charged monomers form ion-pairs too. The ion-pair energy for such ions is $-e^2/4\pi \epsilon_0 \epsilon_l d$, where $d$ is the dipole length and $\epsilon_l$ is the local dielectric constant in the vicinity of the dipole for the respective set of ion-pairs. There are three types of ion-pairs in an intermediate state - negatively charged monomer-positive counterion, positively charged monomer-negative counterion, and oppositely charged monomers. This results in three sets of $d$ and $\epsilon_l$. 
By counting the number of ion pairs, accordingly, for the two dangling chain parts and the complexed part in the middle, the total electrostatic free energy of the ion-pairs for the entire complex takes the form

\begin{equation}
\frac{F_4}{k_B T}=-\left[2 \delta (N-n) (1-f)+n\delta_{12}\right]\tilde{l}_B,
\label{free4}
\end{equation}
where $\delta$ is the local dielectric mismatch for the backbone of the two dangling polyelectrolyte chains (taken to be equal, $\delta_1=\delta_2 \equiv \delta$), and $\delta_{12}$ is the relevant parameter for the complexed part. $\delta=(\epsilon/\epsilon_l)(l/d)$ (in general, it can have three different values for our model), which denotes the disparity between the local ($\epsilon_l$) and bulk ($\epsilon$) dielectric constants of the medium, is defined as usual. Though $\delta$ is commonly termed as the dielecetric mismatch parameter, it takes into account of both the disparity in the bulk to local dielectric permittivity and the measure of the ion size (and, thereby, the local interaction strength for ion pairing).  Equality of $\delta$ for both PEs (along with their length and the size of the monomers) ensures the symmetry that implies $f$ as well as the resulting $R_g$ would be the same for PA and PC. We note that this bound pair energy is proportional to the product of $\delta \tilde{l}_B$ which is termed as the 'electrostatic' or 'Coulomb strength' of the system. Since our work majorly considers a constant $\delta$, the electrostatic strength will mainly be represented by the Bjerrum length $\tilde{l}_B$, which depends on temperature and the dielectric constant of the solvent. The Coulomb strength also significantly modulates the screening effect due to salt, although it does not depend on the salt concentration (which sets the ionic strength of the solution). For the entire work we have set $\delta_{12}=\delta$.

The interaction and entropic free energy arising from the conformational degrees of freedom of different parts of the complex is written in terms of 
the single chain free energy\cite{edwards1979,muthu2004,arindam2010} obtained from Edwards Hamiltonian. For a single charged chain in uniform expansion model, the expression is given by an well-known form
\begin{equation}
\frac{3}{2}[\tilde{l}_1-1-\log \tilde{l}_1]+\frac{4}{3} \left(\frac{3}{2 \pi}\right)^{3/2} w \frac{N^{1/2}}{\tilde{l}_1^{3/2}}+2 \sqrt{\frac{6}{\pi}} f^2 \tilde{l}_B \frac{N^{3/2}}{\tilde{l}_1^{1/2}} \Theta_0(a),
\label{free5Edward}
\end{equation}
where $\tilde{l}_1$ is the dimensionless expansion factor and $w$ the excluded volume parameter mentioned previously. Further, $\langle R^2 \rangle=Nll_1 \equiv Nl^2\tilde{l}_1=6R_g^2$.
The crossover function $\Theta_0(a)$ is defined as,
\begin{equation}
\begin{aligned}
\Theta_0(a) &= \frac{\sqrt{\pi}}{2}\left(\frac{2}{a^{5/2}}-\frac{1}{a^{3/2}}\right)\exp(a)\mathrm{erfc}(\sqrt{a})\\
& +\frac{1}{3a}+\frac{2}{a^2}
-\frac{\sqrt{\pi}}{a^{5/2}}-\frac{\sqrt{\pi}}{2 a^{3/2}},
\end{aligned}
\label{Theta}
\end{equation} 
where $a \equiv \tilde{\kappa}^2 N \tilde{l}_1/6$. The first term in Eq. \ref{free5Edward} is the entropic free energy for conformational degrees of freedom of the PE chain, and the second and third terms arise, respectively, due to excluded volume and electrostatic interactions among the monomers.  We note that the number of monomers in the dangling parts of both the chains is $N-n$ and in the complexed part is $n$. As discussed before,
due to the equality in $w$ and $\delta$ for the prevailing symmetry, $f$ and $\tilde{l}_1$ are the same for PA and PC.
By further noting that the complexed part is uncharged, the conformational free energy of the polymers, including the entropic and the interactive parts,  can be written as
\begin{equation}
\begin{aligned}
\frac{F_5}{k_B T} &=\frac{3}{2} \left[ 2 (\tilde{l}_1-1-\log \tilde{l}_1) + (\tilde{l}_{13}-1-\log \tilde{l}_{13})\right]\\
&+\frac{4}{3} \left(\frac{3}{2 \pi}\right)^{3/2}\left[2 w \frac{(N-n)^{1/2}}{\tilde{l}_1^{3/2}} + w_{12} \frac{n^{1/2}}{\tilde{l}_{13}^{3/2}}\right]\\
& +4 \sqrt{\frac{6}{\pi}} f^2 \tilde{l}_B \frac{(N-n)^{3/2}}{\tilde{l}_1^{1/2}} \Theta_0(a), \\
\end{aligned}
\label{free5}
\end{equation}    
where $w,w_{12}$ are the excluded volume parameters and $\tilde{l}_1, \tilde{l}_{13}$ are the expansion factors, respectively, for the dangling chains and neutral complexed part, $\Theta_0(a)$
is given by Eq. \ref{Theta}, with $a \equiv \tilde{\kappa}^2 (N-n) 
\tilde{l}_1/6$, corresponding to the dangling parts, and $\tilde{\kappa}$ is given by Eq. \ref{kappa}. The first, second, and
third terms in Eq. \ref{free5} are termed $F_{51}/k_B T$ (configurational entropic), 
$F_{52}/k_B T$ (excluded volume), and 
$F_{53}/k_B T$ (electrostatic) for later convenience. In addition,
\begin{equation}
\tilde{R}_g=R_g/l=\sqrt{(N-n)\tilde{l}_1/6} \qquad \mbox{and} \qquad 
\tilde{R}_{g12}=R_{g12}/l=\sqrt{n\tilde{l}_{13}/6}
\label{Rgs}
\end{equation}
are, respectively, the dimensionless radii of gyration of the dangling parts and the
complexed part. Although the electrostatic free energy in Eq. \ref{free5} seems to go like $f^2/\tilde{l}_1^{1/2}$, that is like $Q^2/R_g$ (using $R_g^2=N l^2 \tilde{l}_1/6$), where $Q$ is the charge of the chain, the factor $\Theta_0 (a)$ introduces the non-trivial dependency on salt through $\tilde{\kappa}$ and also a dependency on $R_g$
that is different from $R_g^{-1}$.

We note that for a symmetric PE-pair the dipolar attraction may collapse the complex to a globule with $R \sim N^{1/3}$. However, we ignore such contributions in our main results, because, as we shall see later through detailed calculation, the competitive energetics of free counterions and bound ion-pairs will overwhelm all such variations for the parameter range we have explored.

The free energy of electrostatic interaction between two oppositely charged, dangling PE chains 
(the uncomplexed parts) is modeled using a screened Coulomb potential between two charged spheres. The Coulomb energy of interaction of two charged spherical bodies, each of radius $x$  and center-to-center separation $r$, is given by
\begin{equation}
\frac{U(r)}{k_BT}=Z_1Z_2 \tilde{l}_B \left(\frac{e^{\kappa x}}{1+\kappa x}\right)^2 \frac{e^{-\kappa r}}{r},
\label{dlvo-coulomb}
\end{equation}
where $Z_i$ is the total charge of sphere $i$. The effect of screening of the macroion charges is evident in the factor $\left(\frac{e^{\kappa x}}{1+\kappa x}\right)^2$, which arises due to the presence of ionic clouds in the surrounding of the charged spheres. The ionic clouds appear due to electrostatic induction from its surfaces, which thereby modifies the total charge $Z$ to an effective charge $Z_{eff}=Z \left(\frac{e^{\kappa x}}{1+\kappa x}\right)$. 
In our model the dangling macroions are assumed to be spherical, with the dissociated counterion cloud surrounding them. Recasting Eq. \ref{dlvo-coulomb} in terms of specific PE chain parameters,
we may write
\begin{equation}
\frac{F_6}{k_B T}=-(N-n)^2 f^2 \tilde{l}_B \left(\frac{e^{\tilde{\kappa} \tilde{R}_g}}{1+\tilde{\kappa}\tilde{R}_g}\right)^2 \frac{e^{-\tilde{\kappa}\tilde{r}}}{\tilde{r}},
\label{free6}
\end{equation}
where $\tilde{r}=r/l$ is the dimensionless center-to-center separation of the two lateral spheres,
for which the radius is taken as $R_g$. $r$ shall be equal to the sum of the diameter of the middle sphere (complexed part)
and radii of the dangling parts ($r=2R_g+2R_{g12}$). Scaled by the Kuhn length $l$, $\tilde{r}$ reads
\begin{equation}
\tilde{r}=2 \left\lbrace\left[\frac{(N-n)}{6} \tilde{l}_1\right]^{1/2}+\left[\frac{n}{6} \tilde{l}_{13}\right]^{1/2}\right\rbrace,
\label{centertocenter}
\end{equation}
in terms of the dangling chain length $(N-n)$ and the expansion factors $\tilde{l}_1$ and $\tilde{l}_{13}$. We note that the free ion density has been taken to be uniform throughout the solvent. Ignoring the inhomogeneity that may occur in the space close to the charged chains is an approximation that is made in line with the Debye-H\"uckel theory of ionic fluctuations. Therefore, the same Debye screening length is chosen for the entire solvent, including the space near the charged chains.

The total free energy of the system, consisting of the two oppositely charged PE chains, their counterions, added salt ions, and the surround solvent is now given by
\begin{equation}
F_{\mbox{tot}}/k_B T = (F_1+F_2+F_3+F_4+F_5+F_6)/k_B T,
\label{freentotal}
\end{equation}
which is a function of the variables like the size expansion factor $\tilde{l}_1$ for the 
PE chains, that for the complex formed, $\tilde{l}_{13}$, and the degree of ionization 
$f$ for the PE chains. Since the complexed chain is charge-neutral, its degree of ionization vanishes ($f_3=0$) 
and its size $\tilde{l}_{13}$ is self-consistently minimized to a constant value of unity throughout. 
We must caution that the complexed chain is not Gaussian as implied by this assumption. Charge correlations make it a sub-Gaussian globule\cite{borue1988,borue1990,zhaoyang2006,chen2022}. We shall show later that, for modest parameter values, free energies associated with such conformational changes are negligible compared to major electrostatic contributions, and hence ignored in our main calculations. A similar effect is observed for a negative $w$. Therefore, $w, w_{12}$ are taken to be zero, to keep the number of relevant parameters in the study to a minimum.

\section{Results}

In polyelectrolytes, especially in strong polyelectrolytes, it has been observed that
the free counterion entropy (similar to $F_2$, Eq. \ref{free2}) and the electrostatic energy of ion-pairs arising due to 
condensed counterions (similar to $F_4$, Eq. \ref{free4}) are two overwhelmingly dominant quantities which determine the equilibrium\cite{muthu2004,arindam2010,mitra2017,kudlay2004, fu2016,manning1978}.
Indeed there are other contributions from charge fluctuations, conformational entropy
of flexible polymers, or screened Coulomb interaction among charged species in the system,
but they are found to be at least one order of magnitude or more less than the above
two dominant contributions in most moderate conditions (and low salts)\cite{arindam2010}. The main conjecture of our work is to propose that for this complex, multicomponent system of complexation of two oppositely charged polyions, the same two contributions will continue to dominate the equilibrium and the driving force (or the opposing force) of complexation. We proceed to derive a closed-form analytical result for the charge of the dangling chains, $f$, based on the above
principle.

\subsection{Charge of an Expanded chain - Analytical result}

It can be noted from the free energy that the degree of ionization, $f$, is present in the polymer free energy $F_5$ (Eq. \ref{free5}) only in one term that describes the long ranged, screened electrostatic interactions between charged monomers ($F_{53}$). However, in extended configurations of the chains, the adsorption energy $F_4$ due to the short-range ion pair interactions is more dominant than $F_5$, and shows a greater variation with $f$ in most conditions. For salty solutions, in low salt both the linear and logarithmic terms dominate over the $f^{3/2}$ dependance present in the Debye-H\"uckel fluctuation terms, $F_3$, and in high salt the variation of $F_3$ with $f$ is negligible. With these taken into account one can impose the \textit{adiabatic approximation} for the expanded chain where the configurational part ($F_5$) 
is decoupled from the counterion parts of the total free energy and 
\begin{equation}
F_{\mbox{tot-anal}}=F_{1}+F_{2}+F_{4},
\label{totalfreenanalytical}
\end{equation}
(Eqs. \ref{free1}, \ref{free2}, \ref{free4}, respectively, for the three terms in 
$F_{\mbox{tot-anal}}$) turns out to be the relevant contributions\cite{arindam2010} that determine the charge ($f$) in the expanded chain.
This truncated total free energy will be called the significant part of the free energy from now on. 

This free energy is thus devoid of the expansion factor (as a result of decoupling the configurational part of the free energy) and can be minimised over the single variable $f$ to determine the adiabatic charge in expanded chains. This turns out to be a quadratic equation in $f$

\begin{equation}
f^{2}+\left(\frac{n}{N-n}+\frac{\tilde{c}_{s}}{\tilde{\rho}}+\frac{\exp \left(-\delta \tilde{l}_{B}\right)}{\tilde{\rho}}\right) f-\frac{\exp \left(-\delta \tilde{l}_{B}\right)}{\tilde{\rho}}=0,
\label{chargeanalyticaleqn}
\end{equation}
which gives a closed-form solution for the charge $f$, given by
\begin{equation}
\begin{aligned}
f =&-\frac{1}{2 \tilde{\rho}}\left[\frac{\tilde{\rho} n}{N-n}+\tilde{c}_{s}+\exp \left(-\delta \tilde{l}_{B}\right)\right] \\
+&\frac{1}{2 \tilde{\rho}} \sqrt{\left(\frac{\tilde{\rho} n}{N-n}+\tilde{c}_{s}+\exp \left(-\delta \tilde{l}_{B}\right)\right)^{2}+4 \tilde{\rho} \exp \left(-\delta \tilde{l}_{B}\right)}.
\label{chargeanalyticaln}
\end{aligned}
\end{equation}
In terms of the overlap parameter $\lambda=n / N$, it can be recast as the following that the degree of ionization,
\begin{equation}
\begin{aligned}
f =&-\frac{1}{2 \tilde{\rho}}\left[\frac{\lambda \tilde{\rho}}{1-\lambda}+\tilde{c}_{s}+\exp \left(-\delta \tilde{l}_{B}\right)\right] \\
+&\frac{1}{2 \tilde{\rho}} \sqrt{\left(\frac{\lambda \tilde{\rho}}{1-\lambda}+\tilde{c}_{s}+\exp \left(-\delta \tilde{l}_{B}\right)\right)^{2}+4 \tilde{\rho} \exp \left(-\delta \tilde{l}_{B}\right)}.
\label{chargeanalyticallambda}
\end{aligned}
\end{equation}
The above gives a closed-form expression for the charge of the dangling parts of the two PE chains undergoing complexation in their expanded states.

Now, if we plug back the expression or numerical value for effective charge $f$, for any value of $n$ or overlap $\lambda$, to 
the significant part of the free energy $F_{\mbox{tot-anal}}$ (Eq. \ref{totalfreenanalytical},
along with Eqs. \ref{free1}, \ref{free2}, and \ref{free4}), we get
an analytical expression, and numerical values, for $F_{\mbox{tot-anal}}$.
It is our initial conjecture that the significant part, as explained above, and the results obtained from it,
will closely follow those obtained numerically from the full free energy (Eq. \ref{freentotal}). 
The primary aim of this paper is to verify this conjecture, and quantify the roles of the two major contributions, ion-pair enthalpy and free ion entropy.

Even though the goal of this work is to show that the gain in free ion entropy ($F_2$, Eq. \ref{free2}) and the 
gain (loss) in enthalpy due to bound ion-pairs ($F_4$, Eq. \ref{free4}) dominate the equilibrium overwhelmingly, 
and hence dictate the thermodynamic driving force of complexation, there are other entropic and enthalpic 
contributions to the equilibrium of the system and the complexation process. First we note that, in this model, the complexed chain
being charge neutralized is not extended beyond the Gaussian size anymore (actually it may take sub-Gaussian conformations, depending on the Coulomb strength, as we shall see later), which may lend a small flexibility and entropy to the system (the first term in $F_5$, Eq. \ref{free5Edward}) after complexation, depending on its proximity to Gaussian size.
Further, the dangling parts of the PE chains shall interact electrostatically ($F_6$, Eq. \ref{free6}). 
Even within the danglings parts of each individual PE chains, the uncompensated monomers
will electrostatically repel each other (the third term in $F_5$, Eq. \ref{free5Edward}). 
Eventually in this work we shall make a comparative study of these free energies of complexation, which
will establish the dominance of $F_2$ and $F_4$.

As charge interactions typically overwhelm excluded volume interactions, 
especially in good solvents, we choose the latter to be neglible, rendering
$w=0$ and $w_{12}=0$ for this work. Further, $N=1000$ and 
$\Omega/l^3=1000/0.0005=2 \times 10^6$ (a very low monomer density of
$\sim 0.0005$) are chosen. For all the analyses the Coulomb strength 
in the form of Bjerrum length ($\tilde{l}_B$) and dielectric mismatch ($\delta$) are varied. In addition, for salty cases, the salt concentration ($\tilde{c}_s$) is varied within a limited range.

\subsection{Configurational Properties}

\begin{figure}[h]
\centering
\includegraphics[height=3.95cm,width=4.43cm]{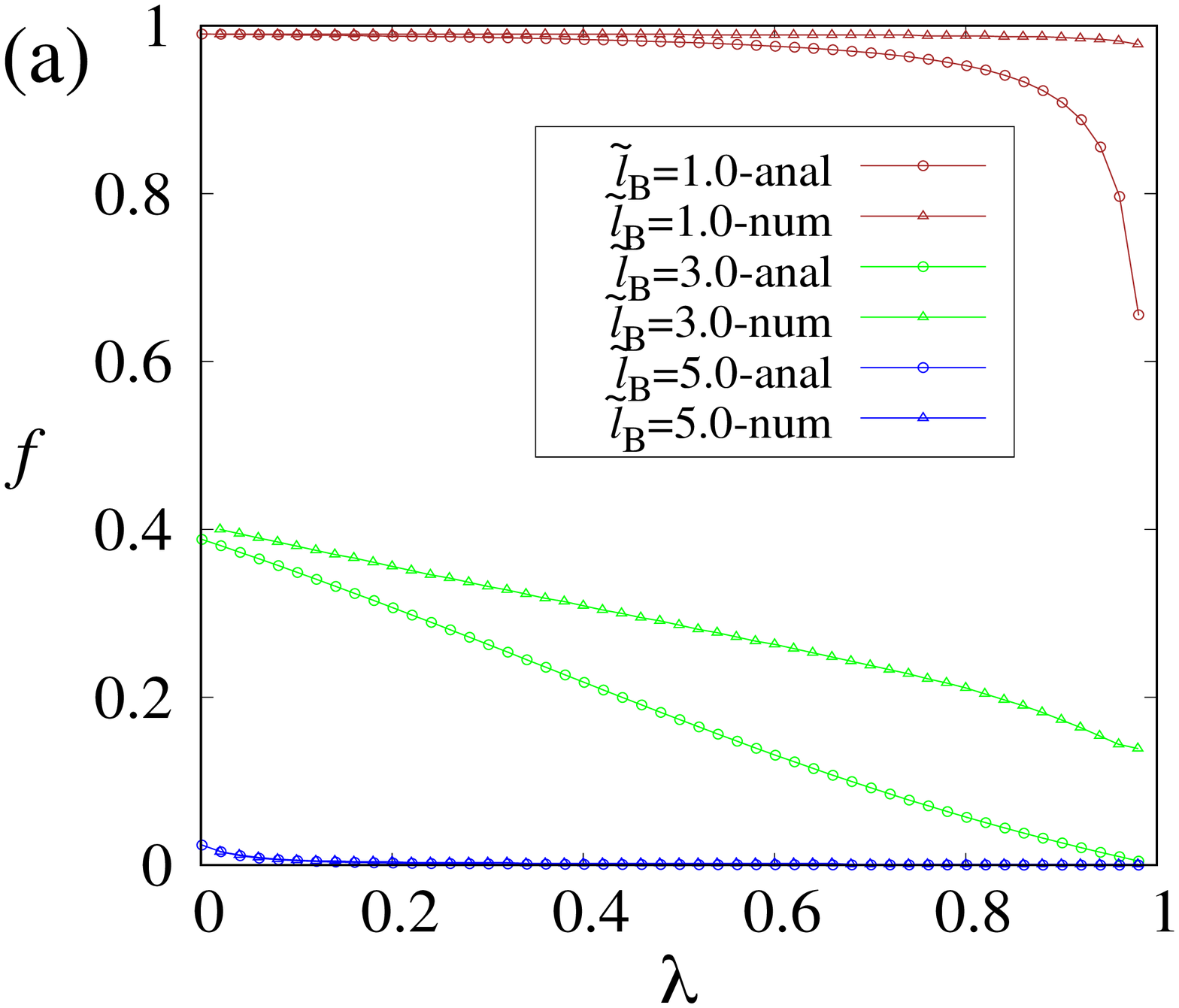}~~
\includegraphics[height=3.95cm,width=4.4cm]{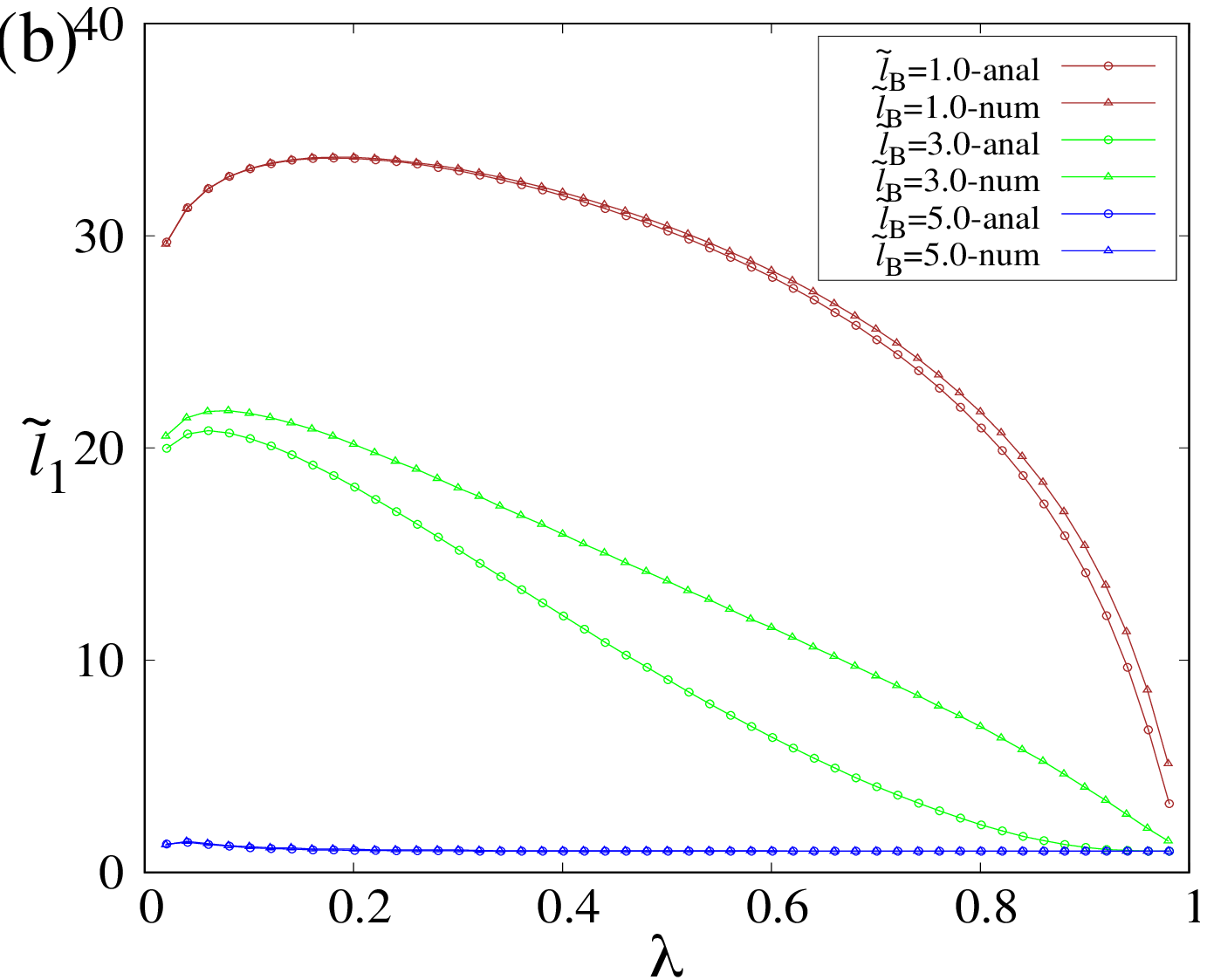}\\
\includegraphics[height=3.95cm,width=4.9cm]{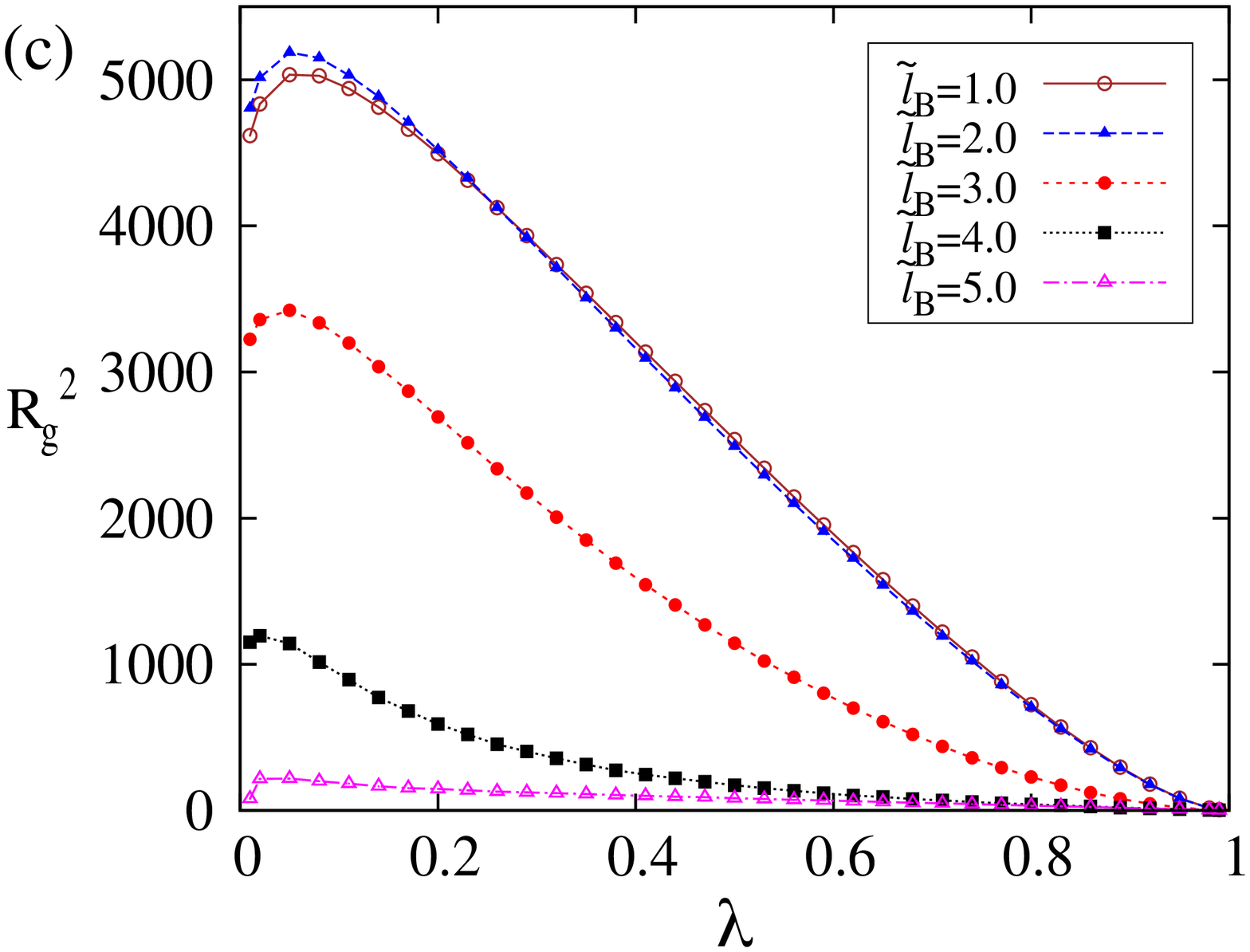} 
\caption{Analytical result for (a) variation of the charge, $f$,
(b) the size, $\tilde{l}_1$, of the dangling, uncomplexed parts of the oppositely charge PE chains, with overlap $\lambda$. The full numerical solution is presented along with for comparison. Parameters taken were: $\delta=3.0$, $\tilde{c}_s=0.0$ and $\tilde{l}_B=1,3,5$. (c) the radius of gyration squared ($R_g^2$) of the dangling, uncomplexed parts of the oppositely charge PE chains, with overlap $\lambda$ is presented only from the full numerical scheme, for $\delta=3.0$ and $\tilde{l}_B=1,2,3,4,5$.}
\label{chargesizeanal}
\end{figure}

We present the charge ($f$) and size ($\tilde{l}_1$) of the dangling chains at an intermediate step of overlap 
$\lambda$ in Fig. \ref{chargesizeanal}. The full numerical results [obtained by minimizing
$F_{\mbox{tot}}$ (Eq. \ref{freentotal})] and the analytical results are compared. For the latter, the analytical value of $f$ is
taken from Eq. \ref{chargeanalyticaln}, and the value of $\tilde{l}_1$ is obtained by plugging
$f$ in the part of free energy consisting of the size, i.e., $F_5+F_6$ (Eqs. \ref{free5} and \ref{free6}, respectively), and then minimizing 
$F_5+F_6$ with respect to $\tilde{l}_1$ only. The similarity between the full numerical and significant analytical results found in Figs. \ref {chargesizeanal}(a) for charge and Fig. \ref{chargesizeanal}(b) 
for size (expansion factor) is reassuring, and indicates that indeed the free ion entropy ($F_2$) and bound-charge enthalpy ($F_4$) are going to overwhelmingly compete to determine the equilibrium, suppressing all other effects substantially. This we shall verify in the next subsection.

To analyze the results, we note that with increasing values of overlap ($\lambda$), the chains progressively lose out monomers in the mutual adsorption (complexation) process (Fig. \ref{overlap}). 
Fig. \ref{chargesizeanal}(a) shows that the charge (which in other words is the fraction of uncompensated monomers for the
dangling chains, $f$) only moderately decreases with decreasing length of dangling chains, and the trend stays the same for all values of the Bjerrum length $\tilde{l}_B$ (the dielectric parameter $\delta=3$ here). Obviously, for higher $\tilde{l}_B$ (lower temperature) the charge is lower for all values of $\lambda$, as more counterions remain condensed on the
dangling chains. This is a salt-free case with $\tilde{c}_s=0$.

We know that for extended conformations the size of the chain is mostly
dictated by its charge\cite{arindam2010}. Similar trend is visible 
in Fig. \ref{chargesizeanal}(b),(c) as well. The expansion factor (or, size for this fixed $N$) $\tilde{l}_1$
decreases with $\lambda$. 
Fig. \ref{chargesizeanal}(c) shows us the absolute size ($\tilde{R}_g=R_g/l
=\sqrt{(N-n)\tilde{l}_1/6}$), which obviously decreases with $\lambda=n/N$.
$\tilde{l}_1$, the expansion factor, is observed
to decrease with charge, as expected\cite{zhaoyang2006}. One may notice a small initial non-monotonicity in the form of a hump in the variation of $\tilde{l}_1$ 
(as well as $\tilde{R}_g$), which is not expected as the size should monotonically go down for a polymer which is being depleted of monomers. This non-monotonicity can be attributed to the coupling of the two dangling chains (with monomers $N-n$ in each) through the binding electrostatic interaction ($F_6$). Two free chains without any interaction between them gives monotonic decrease in their sizes, but the screened Coulomb part($e^{-\kappa r}/r$) works up a dominant rise in the size in small separation lengths $r$. After small amount of overlap this dies down, as the neutral, complexed chain in the middle grows, and the size of the dangling chains starts reducing monotonically. At high temperatures the thermal
energy overwhelms the electrostatic repulsion among like-charged monomers,
leading to Gaussian chain sizes. The radius of gyration also follows a similar trend to that of $\tilde{l}_1$ 
[Fig. \ref{chargesizeanal}(c)] for all values of the parameters, but experiences a sharper decrease as a result of a quadratic dependence.    

\subsection{Free Energy Landscapes} 

In this process of complexation
of two oppsoitely charged PEs, like in most other charged polymeric systems, the free
energy components and their trends with degree of overlap ($\lambda$) depend
on the entropy of free ions, conformational entropy of chains, and electrostatic interactions. 
Such interactions are modulated both by the Coulomb strength 
($\delta \tilde{l}_B$) and screening of salt ions ($\kappa$, Eq. \ref{kappa}). The total free energy 
landscape as a function of overlap dictates whether the process is uphill or downhill with or 
without an energy barrier. The speed of the process of complexation depends on the slope
of the free energy versus overlap curves.
 
For example, in an one-channel polymer translocation process the chain threads into a pore from one compartment 
and proceeds to another. The process involves a barrier that originates mainly from the pore which acts as a confinement for the chain as it nucleates there before translocating to the other side\cite{park1998,muthu1999}. A natural question arises whether the process of complexation which is an overlap between monomers faces similar energy barriers, and if it does, for which ambient conditions of temperature, 
salt, and dielectricity of the solvent and polymers. Analysis
of individual components of the free energy as functions of overlap would help
us both identify such barriers as well as determine the relative merits of such energy components
in promoting (or opposing) the complexation.  

Therfore, in this section, we quantitatively evaluate the free energy components $F_1-F_6$ (Eq. \ref{free1}, \ref{free2}, \ref{free3}, \ref{free4}, \ref{free5}, \ref{free6}) and the total free energy $F_{\mbox{tot}}$ (Eq. \ref{freentotal}) of
the system of two complexing chains, counterions, salt ions, and solvent as functions of overlap $\lambda=n/N$.
The free energy of complexation is obtained from the difference in the respective quantities at two values - $\lambda=0$ for fully separated
chains and $\lambda=1$ for fully complexed chains. This analysis provides insight into the driving force of
complexation. 

\subsubsection{The analytical result - dominance of free ion entropy and ion-pair energy}

As explained before, the main proposition of this work is to verify the conjecture that free ion entropy
($F_2$) and bound-pair energy ($F_4$) dominate the free energy behaviours, and to quantify their
relative importance. Therefore, we first calculate the significant part of the total free energy, $F_{\mbox{tot-anal}}$ (Eq. \ref{totalfreenanalytical}), from the analytical expressions mentioned before (Eq. \ref{chargeanalyticaln} applied to Eq. \ref{totalfreenanalytical}).

\begin{figure}[h]
\centering
\includegraphics[height=3.95cm,width=4.43cm]{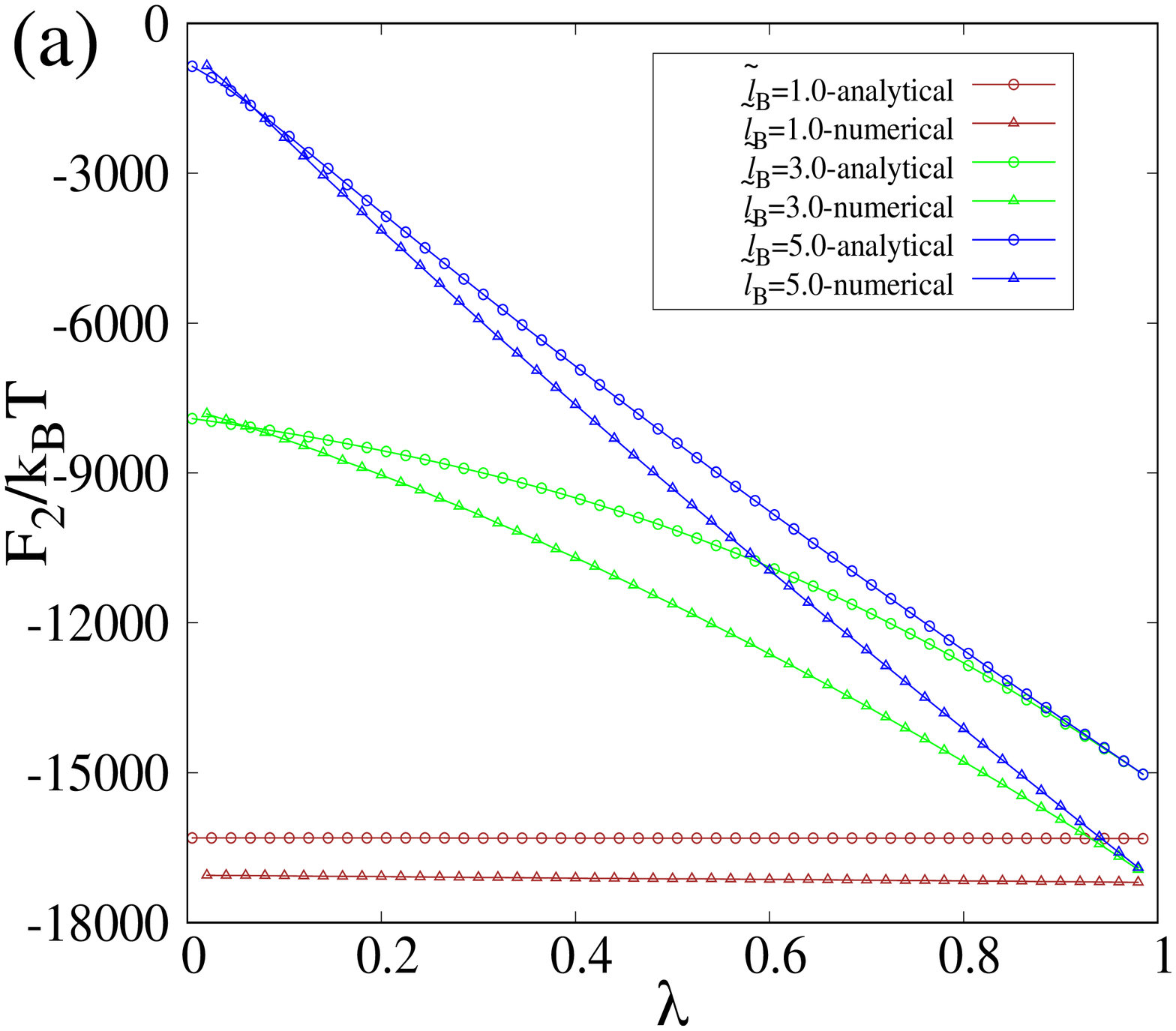}~~
\includegraphics[height=3.95cm,width=4.4cm]{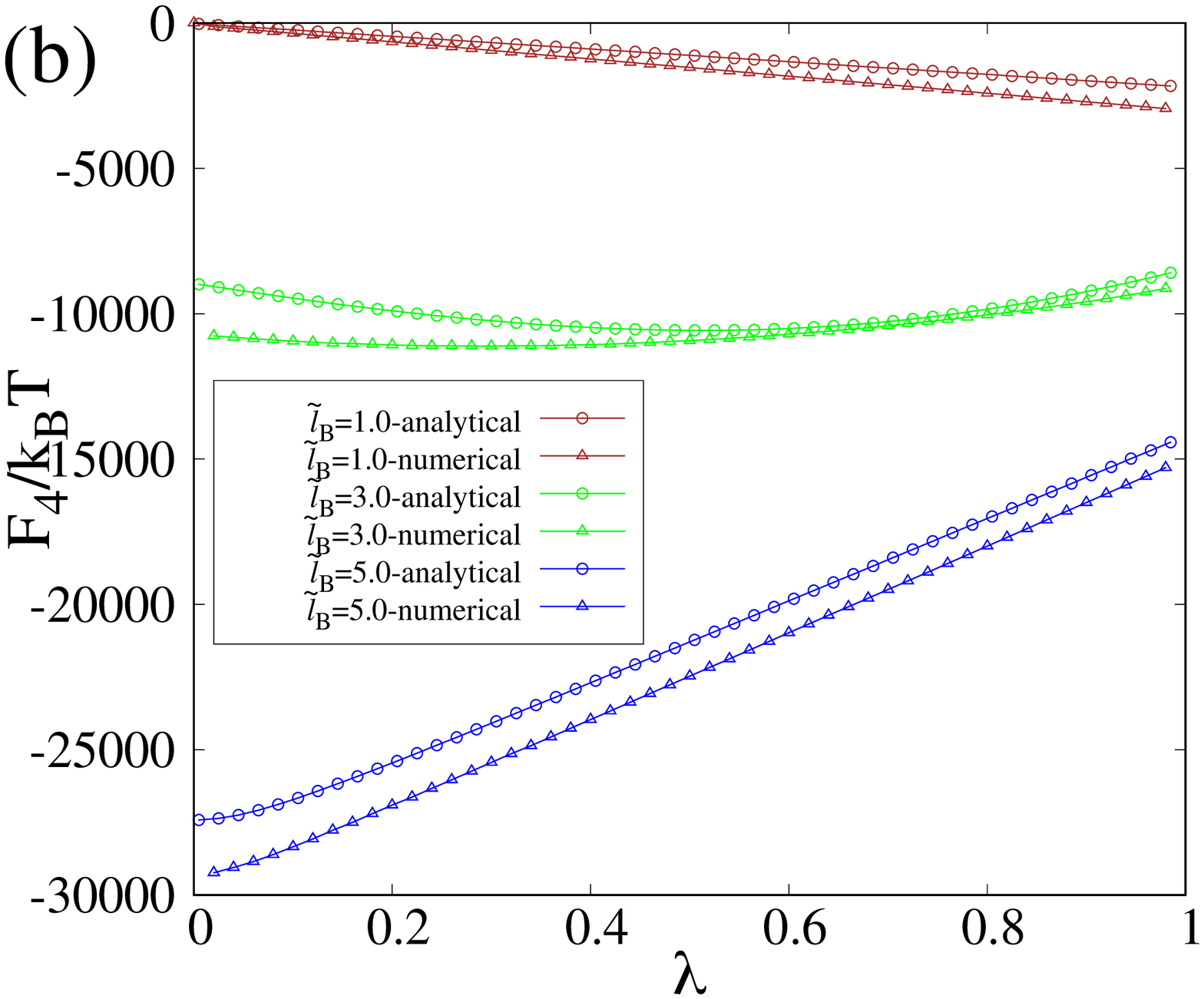}\\
\caption{(a) Analytical result for variation of the free energy, $F_2/k_B T$, due to free ion entropy and
(b) due to ion-pair formation ($F_4/k_B T$) with extent of overlap $\lambda$. The full numerical
solution is presented along with for comparison.Parameters taken were: $\delta=3.0$, $\tilde{c}_s=0.0$, and
$\tilde{l}_B=1,3,5$.}
  \label{free2free4anal}
\end{figure}

\begin{figure}[h]
\centering
\includegraphics[height=6.50cm,width=8.00cm]{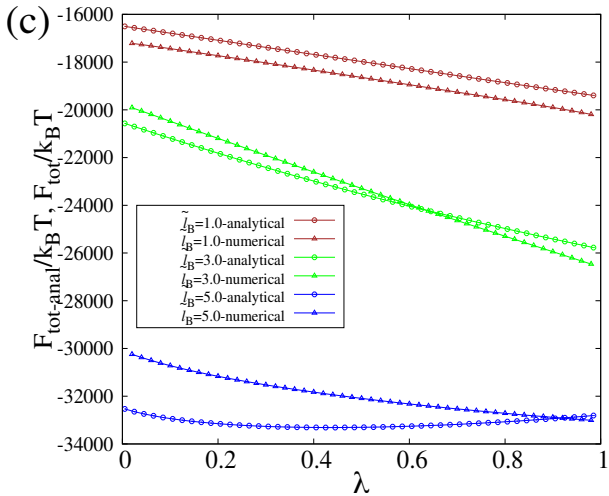}~~
\caption{Analytical result for variation of the significant part of the total free energy, 
$F_{\mbox{tot-anal}}/k_B T = (F_1 + F_2 + F_4)/k_B T$, with extent of overlap $\lambda$. The full numerical
solution is presented along with for comparison.Parameters taken were: $\delta=3.0$, $\tilde{c}_s=0.0,$ and
$\tilde{l}_B=1,3,5$.}
  \label{freetanalfig}
\end{figure}

The results are straightforward, and given in Fig. \ref{free2free4anal}, for the $F_2$ and $F_4$ parts, and in Fig. \ref{freetanalfig}, for the significant part of the total free energy, $F_{\mbox{tot-anal}}$. 
In both
figures the analytical results are presented, along with the full numerical 
solution of the total free energy considering all six terms from $F_1$ to $F_6$ (Eq. \ref{freentotal}). 
Parameters taken were: $\delta=3.0$, $\tilde{c}_s=0.0,$ and $\tilde{l}_B=1,3,5$.

The practically close quantitative similarities between the analytical and numerical results for the total
free energy (Fig. \ref{freetanalfig}) validate the choice of the significant part of the total free energy
to be correct. In other words, if we consider only $F_1, F_2, F_4$, and ignore $F_3, F_5$ (including
$F_{51}, F_{52}, F_{53}$), $F_6$ and do the analysis of energetics, we shall get back all the results with close
quantitative correctness. This vindicates our initial conjecture
that, despite the complexity of the system made of two oppositely charged flexible
polyions, their respective counterion clouds and solvent, only the free ion entropy and the ion-pair
energies are the two factors which overwhelmingly dominate the system equilibrium. One may note that the polymer free energy terms $F_5$ and $F_6$ will be required to determine the size of the dangling chains, although they will not affect the equilibrium significantly.  Therefore, the
dominance of these two effects continue to be observed starting from simple single PE chain 
system\cite{joanny1990,winkler1998,muthu2004,arindam2010,mitra2017}, 
PE solutions\cite{joanny1996,kramarenko2002,hua-arindam2010}, 
PE gels\cite{khokhlov1994,hua-mitra2012,swati2015,swati2017} to more complicated PE systems such as a complexed pair of oppositely charged chains. Fig. \ref{free2free4anal} shows the trends of $F_2$ and
$F_4$ with $\lambda$, for both the analytical and numerical solutions, and the similarities, along with
the trends in $F_{\mbox{tot-anal}}$ and the similarities in orders of mangnitude of $F_2, F_4$ and
$F_{\mbox{tot-anal}}$ (Fig. \ref{freetanalfig}), ensure that indeed $F_2$ and $F_4$ constitute the 
significant majority part of the total free energy. 
Many simulations\cite{winkler2002,hayashi2004,zhaoyang2006,veis2011,semenov2012,dzubiella2016} suggest the same mechanism, and observe a potential of mean force (PMF), that is equivalent to the free energy in our model, decreasing linearly (hence, monotonically) with the reaction coordinate [center-of-mass (COM) distance between the two chains]. More details of this comparison is given in Sec. 'Comparison to Simulations'.  

We shall explain the physics behind the trends observed above in the complete physical description
of all free energy components presented in the next subsection.

\subsubsection{The full numerical results using the total free energy}

After presenting the main proposition of our work, and verifying it analytically, 
in what follows we present the full results for several physical quantities of relevance. All results are derived by numerically
minimizing the total free energy ($F_{\mbox{tot}}$, Eq. \ref{freentotal}) with respect to
charge ($f$) and size ($\tilde{l}_1$) for each value of overlap $\lambda$ of the polyion chains. 

\begin{figure}[h]
\centering
\includegraphics[height=3.95cm,width=4.43cm]{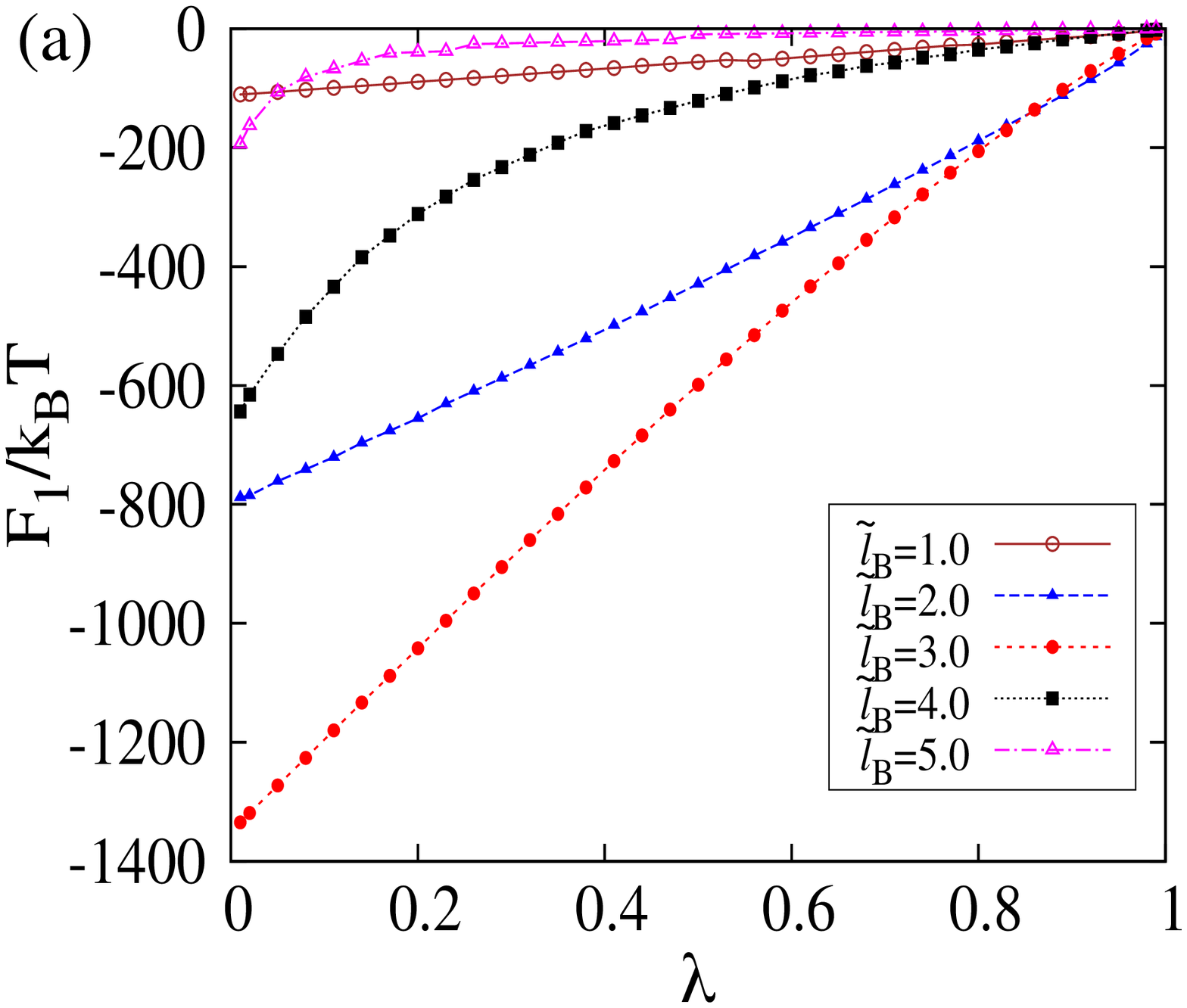}~~
\includegraphics[height=3.95cm,width=4.4cm]{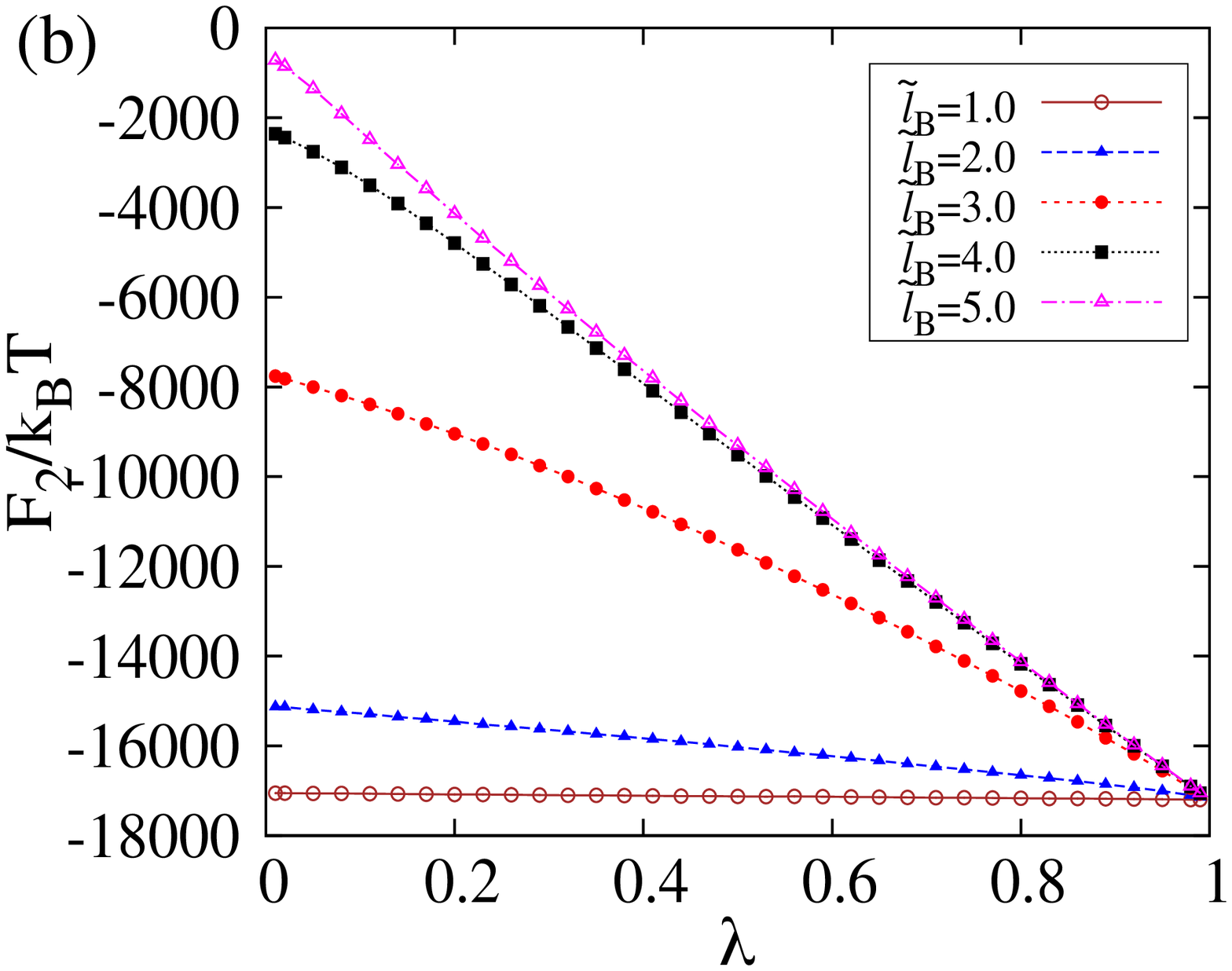}\\
\caption{(a)Variation of free energy arising from the (a) translational entropy of condensed counterions ($F_1/k_B T$), (b) entropy of free counterions($F_2/k_B T$) with extent of overlap $\lambda$, for $\tilde{l}_B=1,2,3,4,5$ and $\delta=3.0, \tilde{c}_s=0.0, N=1000$}
  \label{free1free2}
\end{figure}

Fig. \ref{free1free2} (a) shows the trends of the free energy $F_1$ 
(Eq. \ref{free1}), arising from the translational entropy of the counterions condensed on 
dangling chain backbones (for both PA and PC) (Fig. \ref{overlap}).  As expected, $F_1$
(absolute value) decreases with $\lambda$ with depleting number of available
monomers in the dangling chains. However, when we fix the value of $\lambda$, this entropy
is maximum when half of the available monomers in a chain are compensated by counterions. This condition is satisfied for intermediate values of Coulomb
Strength ($\delta \tilde{l}_B \sim 9$). Hence we observe the non-monotonic 
trend in $F_1$ as a function of the Bjerrum length, $\tilde{l}_B$, for a fixed value of $\lambda$.

Fig. \ref{free1free2} (b) shows the trends of the free energy $F_2$ 
(Eq. \ref{free2}), arising from the ideal gas entropy of the released
counterions (and salt ions in the salty case) coming from the polyions. 
These free ions enjoy the available volume to maximize their entropy. 
First, we note that the free ion entropy ($F_2$) is an order of magnitude 
higher than the translational entropy of condensed counterions ($F_1$). 
Second, this free energy gain for a fixed temperature (or $\tilde{l}_B$) monotonically increases both with overlap ($\lambda$) and with $\tilde{l}_B$ for a fixed value of overlap.
For high $\tilde{l}_B$, the Coulomb interaction will be very effective.
Before overlap, when the PC and PA are disengaged ($\lambda=0$),
almost all counterions ($N$ counter-anions and $N$ counter cations, respectively) remain adsorbed on the monomers. After complete overlap, when the chains are
fully complexed ($\lambda=1$), the oppositely charged monomers from the two
chains form $N$ bound-pairs, and $2N$ counterions are released to the solution.
This explains the maximal change in the entropy of released counterions at high
$\tilde{l}_B$. However, for low $\tilde{l}_B$, i.e., for high temperatures,
the counterions do not adsorb (condense) on the chains even before overlap, and
hence there is no gain in entropy, although the absolute value of entropy remains
high throughout the overlap process. For intermediate values of $\tilde{l}_B$,
the chains are only partially ionized (with released counterions) to start with but 
fully ionized after complexation. Therefore, the entropy gain is always positive, and is maximal and saturated for a certain high value of $\tilde{l}_B$.

\begin{figure}[h]
\centering
\includegraphics[height=3.95cm,width=4.43cm]{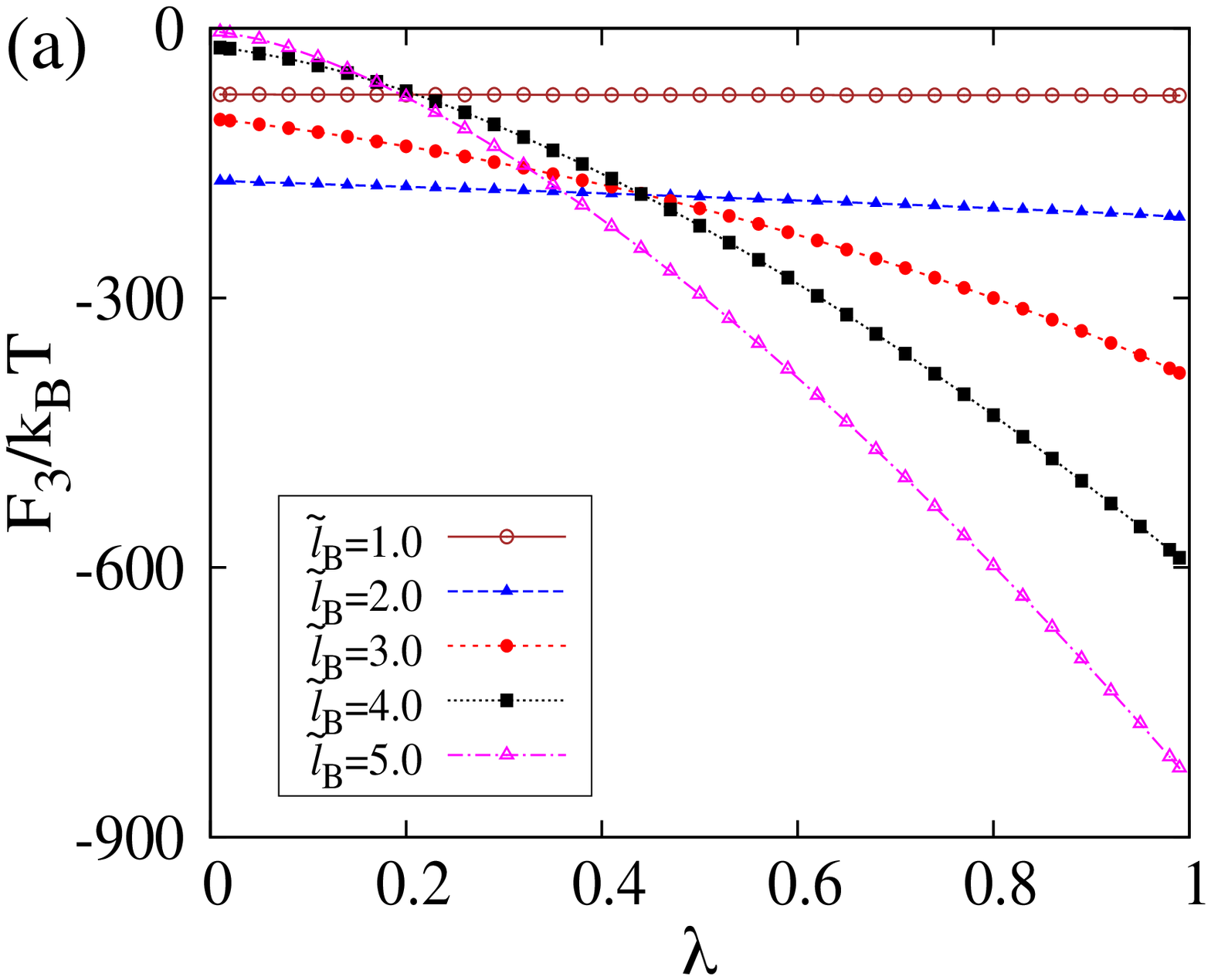}~~
\includegraphics[height=3.95cm,width=4.43cm]{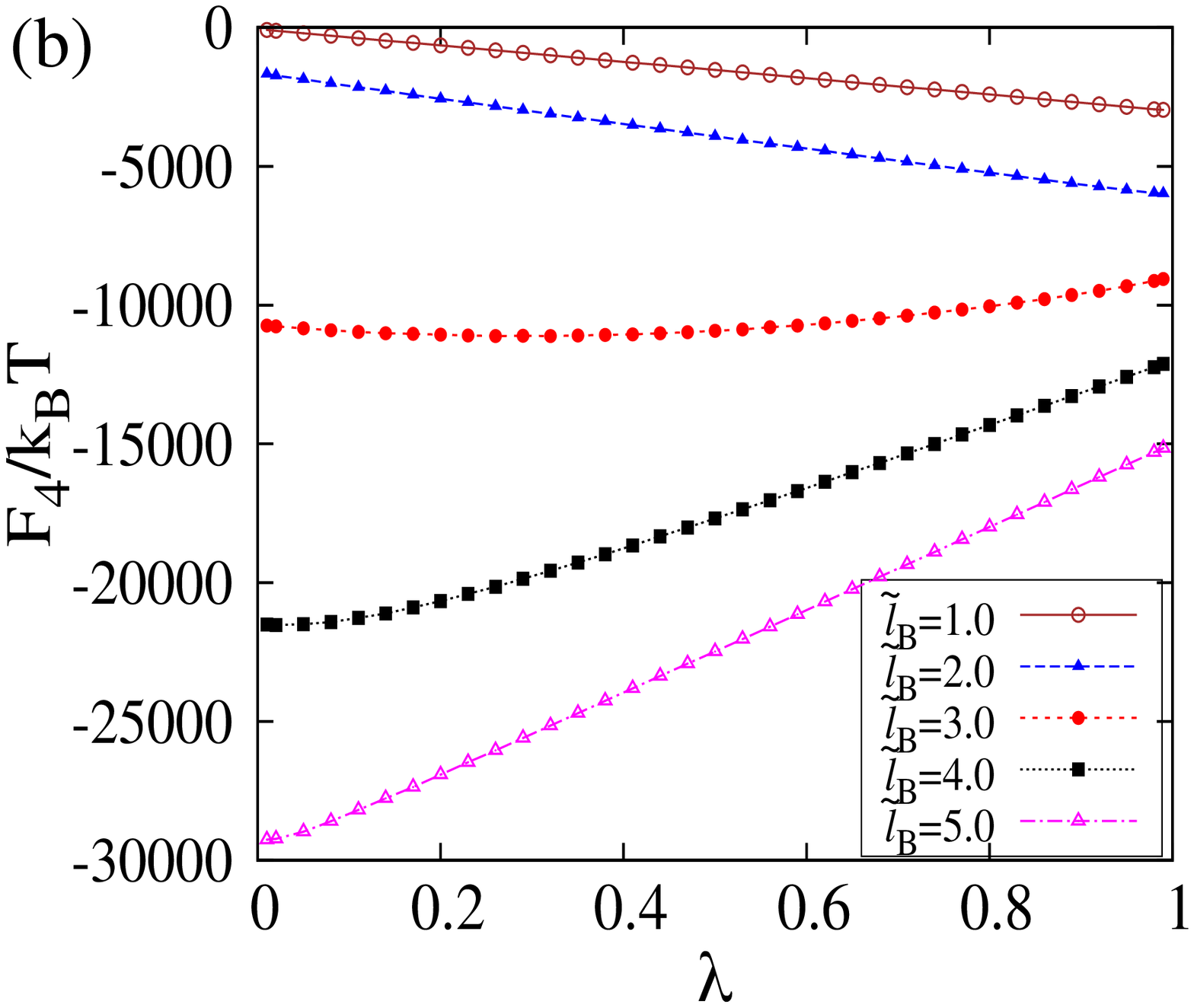}\\
\caption{(a)Variation of (a)free energy of fluctuation of dissociated ions($F_3/k_B T$), (b) enthalpy of bound ion-pairs($F_4/k_B T$) with extent of overlap $\lambda$. Parameters taken were:$\tilde{l}_B=1,2,3,4,5$, $\delta=3.0, \tilde{c}_s=0.0, N=1000$.}
\label{free3free4}
\end{figure}

The free energy contribution $F_3$ (Eq. \ref{free3}) due to the fluctuation in density of free 
ions is not very significant for salt-free conditions (as the free
ion density is low), as shown in Fig. \ref{free3free4}(a). Again, for high
temperatures (low $\tilde{l}_B$s) almost all the counterions stay released 
before and after complexation. The absolute value of $F_3$ is also low for low 
$\tilde{l}_B$. Therefore, both the absolute value and the change in $F_3$ due to complexation is
minimal. For higher values of $\tilde{l}_B$, however, the chains start 
with progressively more adsorbed counterions all of which get released after
complexation. Hence, both the absolute value of $F_3$ and its variation increase with overlap at
lower temperatures. $F_3$, as expected and just like $F_1$, remains one order of magnitude smaller 
than the free ion entropy contribution ($F_2$).

\begin{figure}[h]
\centering
\includegraphics[height=9.0cm,width=6.5cm]{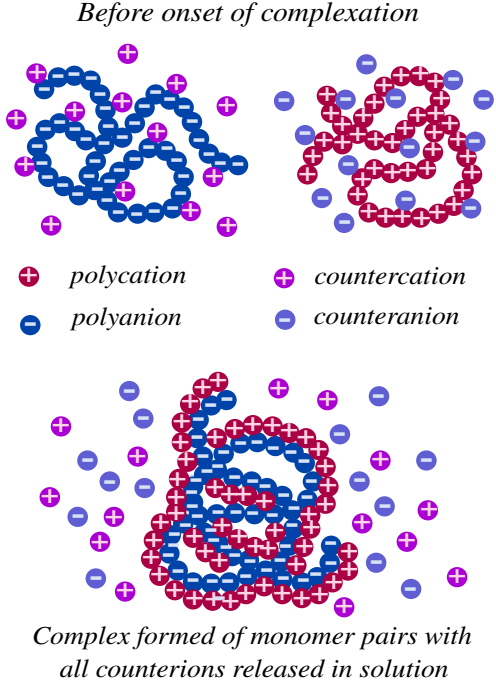}~~~
  \caption{Schematic of the overlapping polyelectrolyte chains forming the neutral complex in between, showing the process of counrerion release.}
  \label{polymer-counterion}
\end{figure}

This brings us to the most important contribution to the free enery $F_4$ (Eq. \ref{free4}),
other than $F_2$ (due to free ion entropy), as shown in Fig. \ref{free3free4}(b). $F_4$ is just the electrostatic 
attractive energy stored in bound ion-pairs (both monomer-monomer in the complex and monomer-counterion in the dangling chains), and is proportional to the number of such pairs. The free ion picture depicted for explaining $F_2$ and $F_3$ (the Debty-H\"uckel term) applies equally well to $F_4$. At low $\tilde{l}_B$s, most ions stay released before and after complexation, resulting in both a very low value and low gain with complexation for free energy component $F_4$. However, it becomes interesting for high $\tilde{l}_B$s. At these low temperatures, before
complexation most counterions are adsorbed to their respective host chains, and
hence the free energy gain due to bound ion-pairs is quite high to start with. However, after complexation, all such counterions are free, but there is only half as many number of bound pairs formed by oppositely charged monomers (see Fig. \ref{polymer-counterion}).
Therefore, there is almost a 50$\%$ drop in electrostatic energy (or enthalpy, as we may call it) during the complexation
process. We can conclude that at low temepartures there is significant enthalpy loss due to complexation of oppositely charged polyions, and {\it the complexation process is actually opposed by the electrostatic
attraction of oppositely charged polyions}! Only once the entropy of free ions is considered in the free energy (or the potential of mean force in simulations) the association of the chains becomes favourable.  At moderate values of $\tilde{l}_B$ (say, 3 here, for which the Coulomb strenth $\delta \tilde{l}_B \sim 9$), there is insignificant change in enthalpy due to bound ion-pairs. At these values of Bjerrum length, before complexation approximately 50$\%$ of the monomers are compensated by counterions for both PA and PC. After complexation, all monomers of both chains form bound pairs leading to conservation of total number of bound ion-pairs in the process. It continues to be remarkable that the enthalpy of bound counterions ($F_4$) is on a far larger scale than those arising from other electrostatic effects ($F_3$ and the part of $F_5$ having the screened Coulomb repulsion among the charged monomers) in solution. From the variations of the entropy gain of free counterions ($F_2$) and enthalpy gain (loss) of ion-pair formation ($F_4$) one may aim to explore whether these two strong, competing effects, especially at moderate or low temperatures, give rise to non-monotonicity in the functional dependence of the free energy on the overlap ($\lambda$) of chains. 

\begin{figure}[h]
\centering
\includegraphics[height=3.9cm,width=4.43cm]{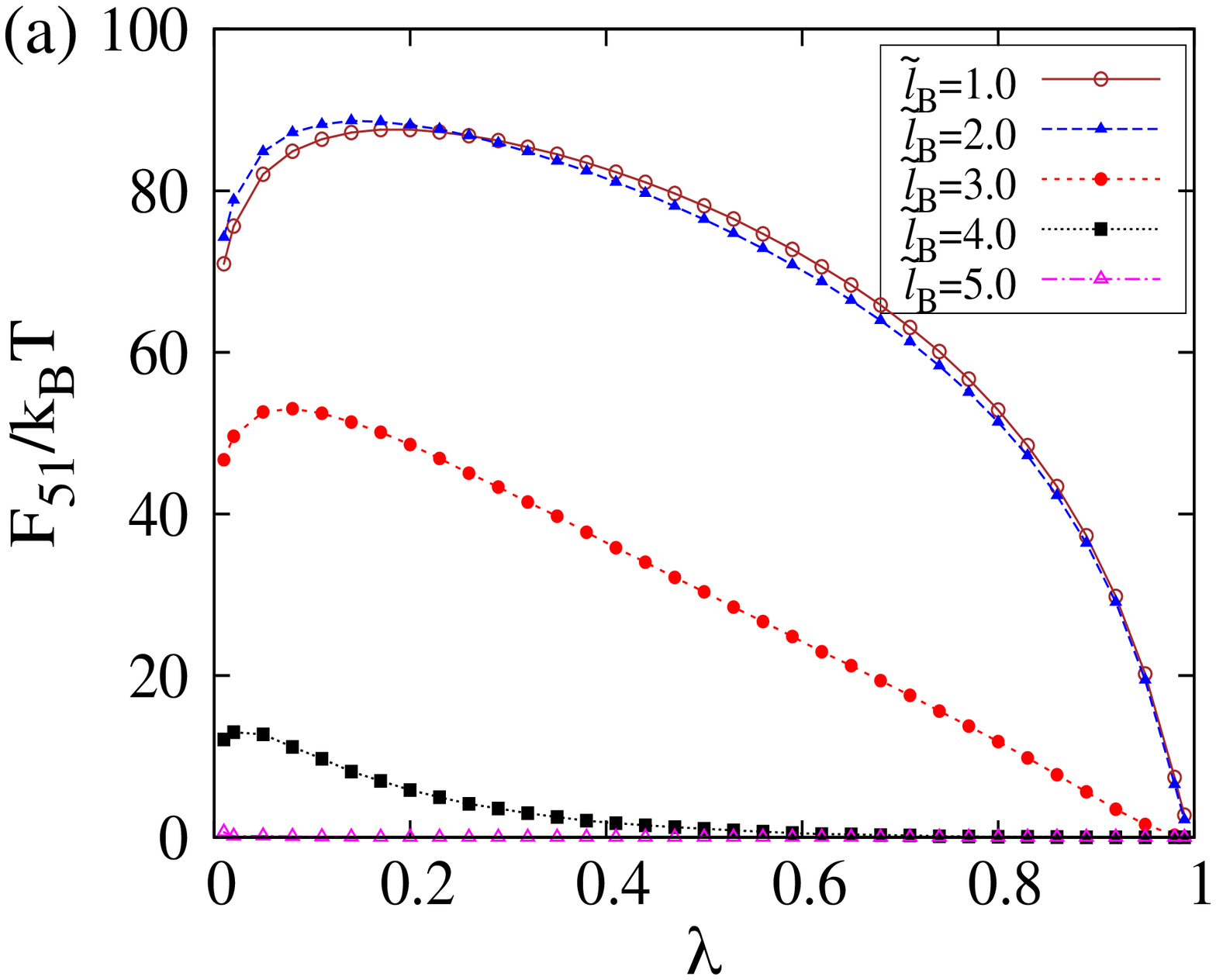}~~
\includegraphics[height=3.9cm,width=4.43cm]{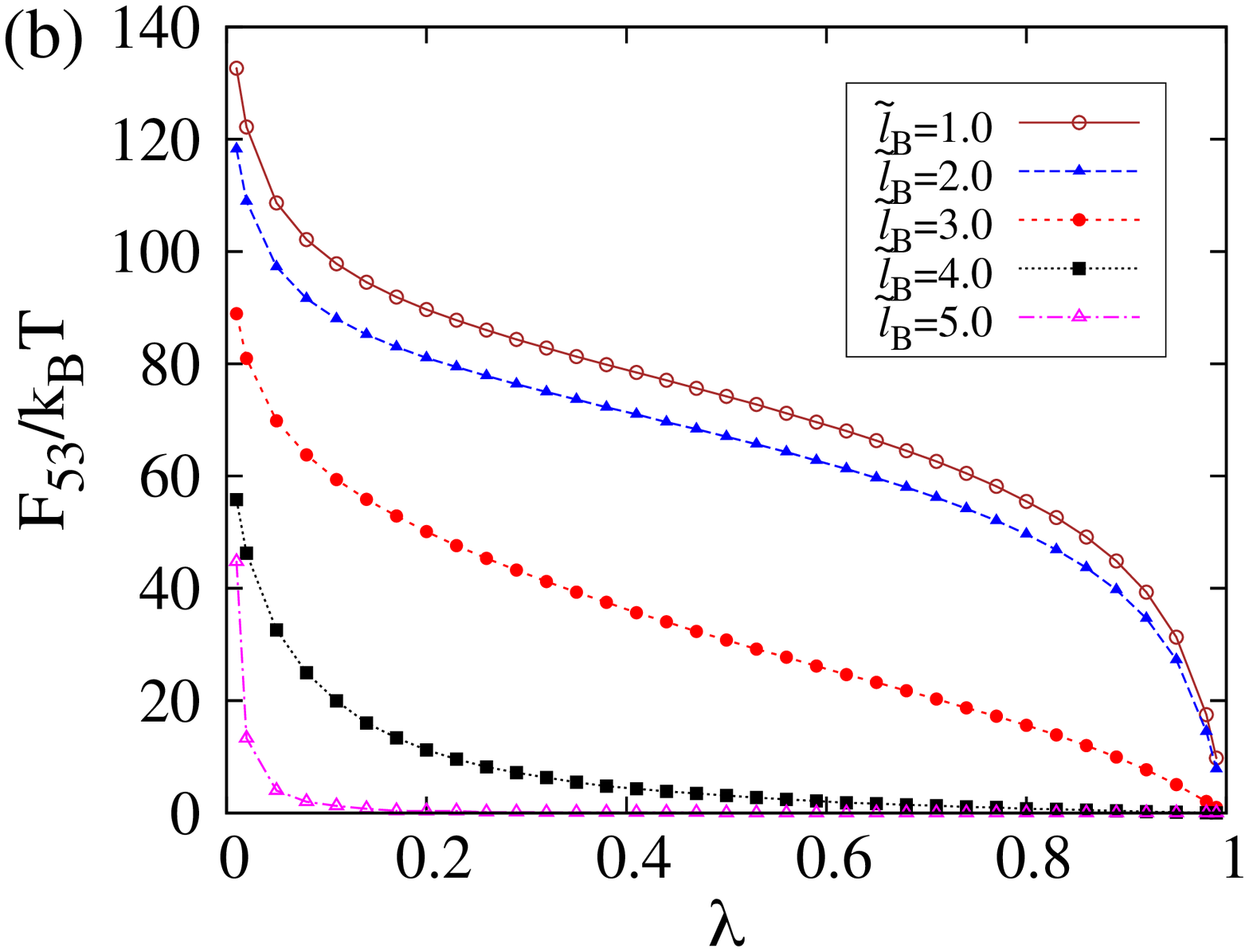}\\
\caption{(a)Variation of the (a)configurational entropy of the dangling, uncomplexed part of the PE chain ($F_{51}/k_B T$), (b) Intra-chain Coulomb repulsion between the segments of the dangling part of the chains ($F_{53}/k_B T$) with extent of overlap $\lambda$. Parameters taken were:$\tilde{l}_B=1,2,3,4,5$, $\delta=3.0, \tilde{c}_s=0.0, N=1000$.}
  \label{free51free53}
\end{figure}

We now shift to the free energy parts corresponding to the configurational properties of the polymer chains. Fig. \ref{free51free53} shows the free energy of the dangling polyion
chains, $F_5$ (Eq. \ref{free5}), for the intra-chain interactions. We note that the term corresponding to the excluded volume interactions (the second term in $F_5$) is zero, because $w=0=w_{12}$. The free energy of the complexed
part of the two chains (the intervening sphere in Fig. \ref{interactions}), arising from the conformational entropy of the monomers (the fourth term in $F_5$) is zero, because in the absence of excluded volume and electrostatic interaction the bound-pairs are assumed to form a freely-jointed random-walk chain, for which the expansion factor is taken as $\tilde{l}_{13}=1$, the Gaussian chain value (we shall see later that charge-correlations make the parts of the chains in the complex sub-Gaussian, but energies related to such conformational changes are very low).  $F_{51}$, the first term in $F_5$ corresponding to the conformational entropy of the dangling chain parts, approximately goes as three times the size 
expansion factor of the chain $\tilde{l}_1$, which is reflected in Fig. \ref{free51free53}(a). $F_{53}$, the third term in $F_5$ 
corresponding to the electrostatic repulsion of uncompensated like-charged 
monomers (for the dangling parts, individually, of both chains), decreases sharply with the degree of overlap 
($\lambda$), because both the degree of ionization, $f$, (nominally) and number
of available monomers (linearly) decrease with $\lambda$. As expected, $F_{53}$ is
progressively lower for higher $\tilde{l}_B$s, as substantial adsorption 
of counterions drastically reduces $f$. We must notice,
however, that compared to free ion entropy, $F_2$, and adsorption enthalpy, $F_4$, the intra-chain conformational and electrostatic free energies
($F_5$ as a whole) are insignificant.

\begin{figure}[h]
\centering
\includegraphics[height=3.9cm,width=4.43cm]{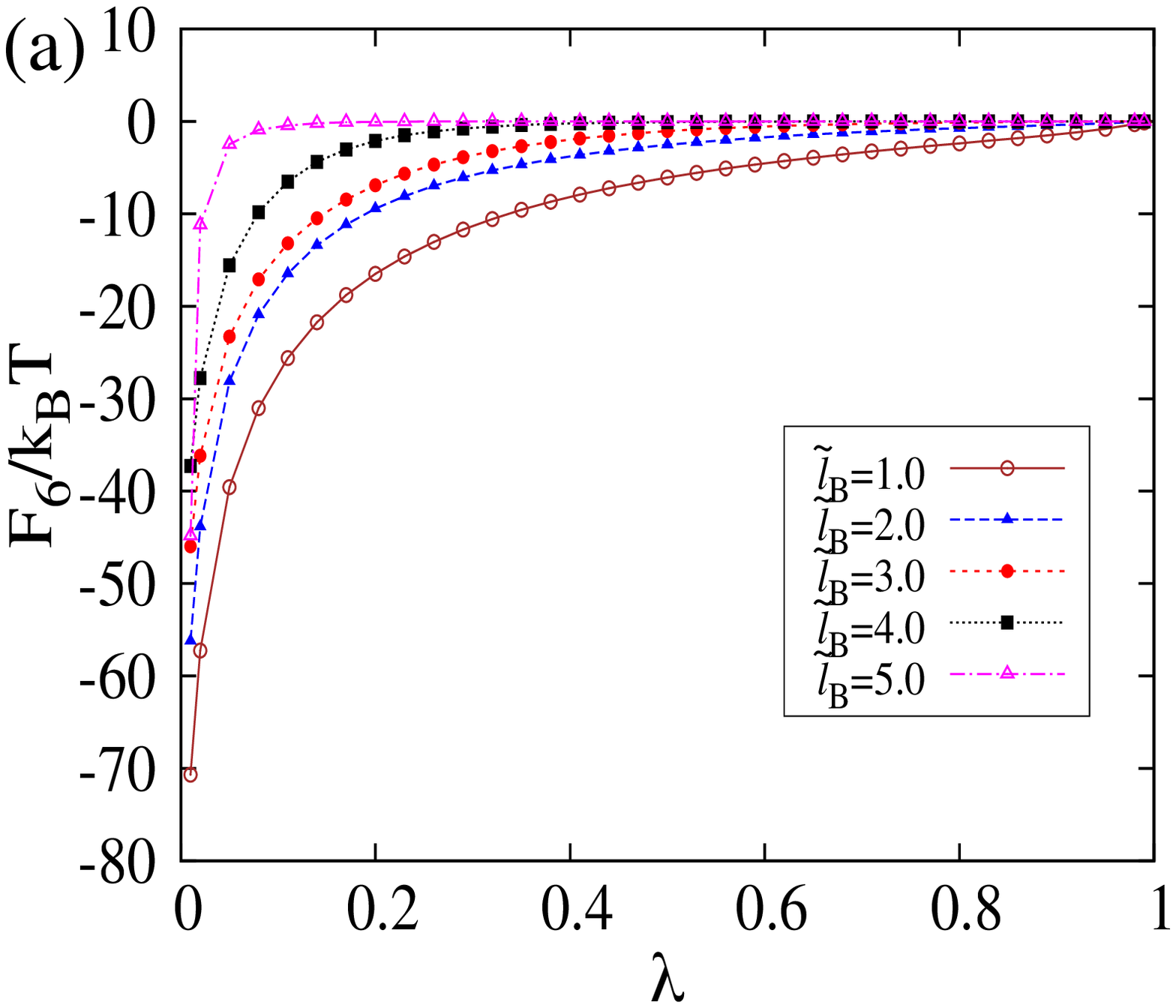}~~
\includegraphics[height=3.9cm,width=4.43cm]{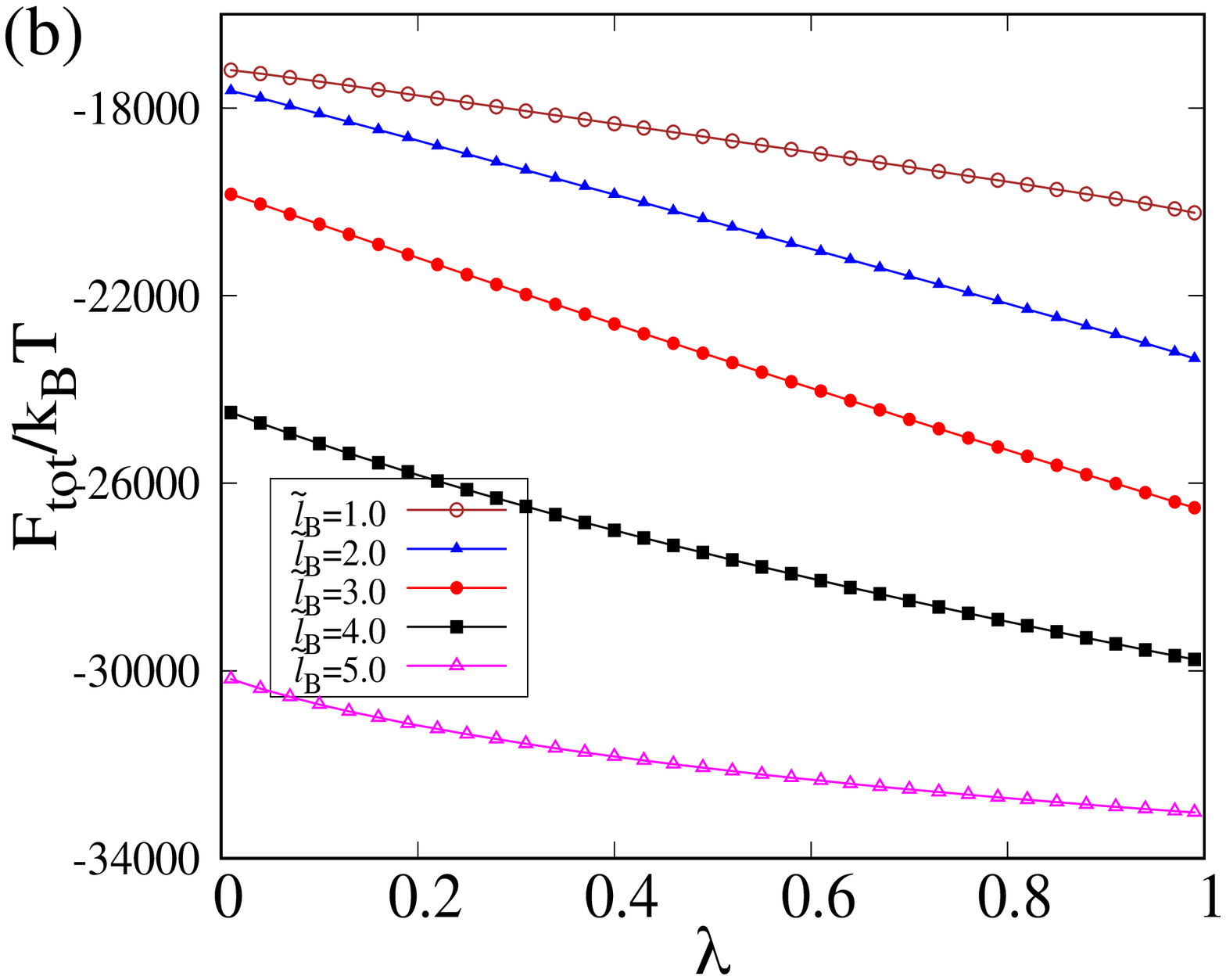}\\
\caption{Variation of (a)mutual Coulomb attraction between segments of the oppositely charged PE chains ($F_{6}/k_B T$), (b) total free energy ($F_{tot}/k_B T$) profile for the two chain-complex system with the extent of overlap. Parameters taken were: $\tilde{l}_B=1,2,3,4,5$, $\delta=3.0, \tilde{c}_s=0.0, N=1000$.}
  \label{free6free_tot}
\end{figure}

Fig. \ref{free6free_tot}(a) shows the inter-chain electrostatic interaction, $F_6$,
between the two dangling chain ends of the PA and PC (Figs. \ref{overlap} and \ref{interactions}). Depletion of monomers from the dangling parts of the respective PEs to enrich the complex reduces the total charge on both of them. We see a rapid drop in the (absolute value) of the interaction
energy with overlap $\lambda$. The reason is two fold. First, the dangling chains lose out monomers 
(hence, the overall charge) progressively with $\lambda$. Second, as the lost monomers form bound ion-pairs 
in the complex that grows in size, it creates a separation between the charged, dangling chains, which interact with screened Coulomb potential that decays much faster than simple $1/r$ potential. $F_6$ versus $\lambda$ varies negative upward and saturates to zero, at which point all the charged monomers from both dangling chains are completely depleted by mutual adsorption (fully complexed state). 

This part of the free energy, although quite miniscule compared to $F_2$ and $F_4$, is worth a discussion on its own merit. Unlike most other electrostatic energy changes, the prominence of
this energy gain is not higher for higher Coulomb strengths (which, as mentioned before, goes as $\tilde{l}_B$ when $\delta$ is fixed). 
One may note that for high temperatures all the counterions will be free even at the start of the complexation, and number of bound pairs, charge, and inverse Debye length take values,  
$n=0, f=1$, and $\kappa \rightarrow 0$ (as $\tilde{l}_B \rightarrow 0$), respectively. The size of the complex $R_{g12}$ would be zero and the separation $r=2R_g$. In that case, $F_6/k_B T$
would simply go as $-N^2 \tilde{l}_B/(2 R_g)$, which is low for low values of $\tilde{l}_B$ ($\rightarrow 0$ at high $T$). 
If one increases $\tilde{l}_B$, there are three different effects on this interaction. 
The charge $f$ goes down with condensation of counterions, the interaction strength
increases linearly with $\tilde{l}_B$ (but for a fixed value of $f$), and
the screening effect of Coulombic interaction between the charged spheres increases.
These conflicting effects lead to a non-monotonic, and counterintuitive, dependency of $F_6$ on
$\tilde{l}_B$ for any fixed value of overlap $\lambda$. In Fig. \ref{free6free_tot}(a) we present only the
range of $\tilde{l}_B$ for which $F_6$ (absolute value) decreases with 
$\tilde{l}_B$. This implies that for this range the charge modulation by condensation takes
a greater role. However, in the context of the complexation drive, the most 
noteworthy fact remains that $F_6$, the energy gain due to this attractive electrostatic interaction 
between the dangling parts of the PC and PA, is two orders of magnitude lower than the 
energy gain due to bound ion-pair formation ($F_4$). 

A comment on ignoring the inhomogeneity of the ion cloud near the charged chains in the calculation of $F_6$ is in order here. As mentioned before, Debye-H\"uckel theory approximates by considering a uniform Debye screening length for the entire solution. This simplification turns out to be justified, in our opinion, by the fact that this term ($F_6$) is quite insignificant in comparison to the entropic and enthalpic free energies in our model. Inclusion of an inverse Debye length $\kappa$ that depends on the inhomogeneities of the ion-cloud will indeed give us a marginally better result, but will make a relatively insignificant part of the free energy appear more complicated. We must admit, however, that any study that focuses on the electrostatic interaction between the uncomplexed parts of the chains must consider the inhomogeneity of this ion cloud.

This brings us to the numerical plot of the total free energy arising out of all terms above
($F_{\mbox{tot}}=F_1+F_2+F_3+F_4+F_5+F_6$, Eq. \ref{freentotal}) in 
Fig. \ref{free6free_tot}(b) (a part if this
result is already presented in the plot of comparison with the analytical result - Fig. \ref{freetanalfig}).
It is evident from the values plotted for all free energy components that free
ion entropy ($F_2$) and ion-pair enthalpy ($F_4$) overwhelm other components
to dominate the total free energy profile. They are of order tens of $k_B T$
per monomer (we note, the degree of polymerization $N=1000$), compared to a fraction of a $k_B T$ per monomer for other components, and this dominance is valid for all values of the overlap. For low $\tilde{l}_B$ (high temperatures), most ions are free before and after
complexation, and the entropy of complexation changes nominally. The driving
force solely arises from the enthalpy of complexation due to ion-pair
formation of oppositely charged monomers. However, as $\tilde{l}_B$ is low, 
the gain from this enthalpy change (proportional to $\tilde{l}_B$) is low as well. For high
$\tilde{l}_B$ (low temperatures), on the other hand, the polyions start with all counterions adsorbed on them, and eventually release all of them once all the
monomers bind to form the complex. Hence, the gain in free-volume entropy of released counterions is maximal. However,
as the complex loses out on almost half the ion pairs, the enthalpy (gain) also
reduces to almost half. Due to these competing effects of ion release and binding, at low temperatures, the enthalpy of
complexation is prohibitive of the process, whereas the entropy of complexation
is strongly supportive, so much so that it overwhelms the enthalpy loss
and drives the process handsomely. This is another counterintuitive result, because
one would expect the electrostatic enthalpy to be dominant over thermal
randomness at low temperatures. For the given set of parameters, which 
are modest values as per previous literature (taking an example of $\tilde{l}_B \sim 3$ and
$\delta \sim 3$ for NaPSS in water, once we note that $l_B=0.7$ nm in water at room temperature and $l$, the monomer size, is 0.25 nm for NaPSS)\cite{muthu2004,delacruz2004,lee2009,larson2009,arindam2010}, the middle range of the parameter space explored for $\tilde{l}_B$ and $\delta$ seem to provide the largest free energy drop, hence the largest thermodynamic drive, for complexation. Almost all free energy versus overlap curves are linear downward, that is consistent with simulation results (discussed in Sec. 'Comparison to Simulations').

We analyze the possibility of small ion-pairing at high Coulomb strengths\cite{zhaoyang2006,fisher1993}, which, if present, may affect the energetics by decreasing the enthalpy loss of complexation. In our theory, all ions and monomers are of the same size ($d=l$), and the local dielectric constant ($\epsilon_l$) is applicable to counterion-monomer and monomer-monomer pairs equally ($\delta_1=\delta_2=\delta_{12} \equiv \delta$). However, the counterion-counterion pair is formed in the bulk for which the bulk dielectric constant (80 for water) is applicable. Therefore, the energy gain for one such ion-pair (small ions) would be $\tilde{l}_B k_BT$, which is less favourable in the absence of $\delta$. For one small ion-pair formed the system will lose one freely moving particle (not two), leading to entropy loss, $\log(\tilde{\rho}+\tilde{c}_s)-1$, of 8.5 $k_B T$ (no salt), 7.5 $k_B T$ ($\tilde{c}_s=0.001$), and 5.6 $k_B T$ for $\tilde{c}_s=0.01$ (this concentration is 10-times the highest salt used in our work), which is independent of the Coulomb strength, and valid for these dilutions (in our case, 100 mM of monomers). Unless $\tilde{l}_B$ reaches 8.5 (no salt) and 7.5 ($\tilde{c}_s=0.001$), we expect no small ion-pairing, the favourability of which further decreases with lower $\tilde{l}_B$'s, while keeping $\delta \tilde{l}_B$ the same. Note that the complex may be unstable at a lower $\tilde{l}_B$, if the enthalpy loss is
enhanced due to a lower affinity of monomers (effectively a lower $\delta_{12}$, caused by larger monomer sizes or the charge distribution). A full numerical minimization, by allowing for a possibility of such ion-pairing and modification of Eqs. \ref{free1} to \ref{freentotal}, confirms their absence, lending credence to our original assumption.  Small ion-pairing will be a progressively important issue for high Coulomb strengths at even higher salt concentrations, a good part of which is beyond the reach of our theory in current form, but needs a detailed look in future.  Small ion-pairing, if present, needs to be analyzed by a modified Debye-H\"uckel theory considering electrostatic monopole-dipole interactions\cite{fisher1993} which may lead to phase separations. For our problem, the critical values for such phase diagrams are  $\rho_c^\star \equiv \rho_c l^3 \equiv \tilde{c}_s=0.030$, that gives $\rho_c$ around 3M for NaPSS and NaCl [the size of Na+ ions is taken around 0.25 nm (the van der Waals radius)], and $T_c^\star=0.06$ that gives $T_c=T_c^\star e^2/4 \pi \epsilon_0 \epsilon k_B l =$50 K (or $\tilde{l}_{Bc} \times 0.06=1$ that leads to $T_c \simeq 55$K, considering that $\tilde{l}_B=3$ corresponds to $T=300$K), which is quite low, but can be higher in solvents with low dielectric constants.

One important note on the dependency of this energetics on the polymer density (system dilution) is in order. For very high dilution it is expected that all counterions will remain free before and after complexation, but the number of ion-pairs will rise sharply after. Hence, the complexation will be enthalpy driven for most Coulomb strengths. For denser systems, for which most of the counter-cations and anions remain condensed on the respective monomers before complexation, there is loss of enthalpy and gain in entropy after complexation (see the discussion on the competing effects of $F_2$ and $F_4$ following Fig. \ref{polymer-counterion}). For very dense systems, other components of the free energy may also become important. We had chosen $\Omega$, the system volume, to be 10, 100 times larger and 10, 100 times smaller compared to $\Omega=2 \times 10^6$, the one used for all results in this work, and had examined the free energy components. The main proposition of the work - the dominance of $F_2$ and $F_4$ over other components of the free energy - remains valid for an excellent range of polymer density, except only for very high densities. We omit the detailed results related to this dilution dependency for lack of space, and plan to present them elsewhere.

We may conclude by noting that the primary challenge of our theory has been to identify approximations which are well justified from the perspective of charged polymers, but to simplify the approach to find major energetic contributions to the complexation of two PE chains. Starting from the full thermodynamic picture, all the terms in the free energy are analyzed to identify the dominant contributions, whereas
all other contributions remain less significant. The approximations help us with closed form analytical formulae, verified by the full numerical solutions, which strengthens the conjectures. In contrast to traditional approaches like field theoretic simulations (FTS) or RPA, which are different theoretical schemes applicable for bulk solutions for which electrostatic effects average out, in this work we wanted to identify the dominant effects within a phenomenological free energy of polymer chains, and to give the minimal description of the complexation process that can be directly verified by simulations.

We note that the simulations show \cite{dzubiella2016,zhaoyang2006,semenov2012,winkler2002,hayashi2004,semenov2012,chen2022} that indeed a strong binding is possible between two strong PE chains with all monomers ionizable, leading to the so-called 'ladder' model of complexation. The model is better applicable for energetics than conformations, elucidates the energy contributions clearly and correctly, which is the main focus in our theory, and also offers intermediate states which help estimate the potential of mean force. Even if one considers a 'scrambled-egg' complex, in which the monomers do not bind sequentially but rather randomly, the results will remain very similar, as long as the ion-pair energy is significant \cite{semenov2012}. We further find that the chain overlap can be sequential, but the final ion-pairing can be considered scrambled within the overlapped part, without much effect on the energetics. This justifies our original assumption that instead of mixing randomly the two oppositely charged polyions (the so-called scrambled egg model) 
a choice of a path of gradual overlap (ladder model) would be acceptable for this work. 

The variation of the total free energy as a function of 
overlap ($\lambda$) affects the speed of complexation. The slope of the curve
at each value of $\lambda$ should control the rate of formation of the 
monomer-monomer ion-pairs.

\subsection{Collapse of Chains in Complex}

Coulomb interactions and charge correlations at sufficiently high Bjerrum lengths collapse the polyions in the stoichiometrically charge-neutralized complex\cite{borue1988,borue1990,zhaoyang2006,chen2022,trejoramos2007,shakya2020,whitmer2018,semenov2012}, and even the individual chains\cite{winkler1998,brilliantov1998,liu2002}, to sub-Gaussian sizes, although the energy involved is expected to be low compared to strong interactions\cite{arindam2010,liu2002,winkler1998} (such as ion-pair formation, $F_4$). The energy of the dipolar attractive electrostatic attraction of the ion-pairs, written as the first step to account for charge correlations\cite{winkler1998,liu2002,muthu2004} in addition to counterion condensation, 
\begin{equation}
\frac{F_{54}}{k_{B} T}=\frac{4}{3}\left(\frac{3}{2 \pi}\right)^{3 / 2} w_{1} \delta^{2}\tilde{l}_{B}^{2} \tilde{d}^{6} \left(2 (1-f)^{2}  \frac{\sqrt{N-n}}{\tilde{l}_{1}^{3 / 2}}+ \frac{\sqrt{n}}{\tilde{l}_{13}^{3 / 2}}\right),
\label{F54tot}
\end{equation}
is similar in form to the second term, $F_{52}/k_B T$, in Eq. \ref{free5}, and consists of two terms corresponding to the monomer-counterion pairs in the dangling parts of both the PE chains (in the mean field $1-f$ dipoles per monomer are created in the dangling chains) and the oppositely charged monomer-monomer pairs formed in the neutral complex (all $n$ composite monomers in the complexed part form dipoles), respectively. Here $w_1$ is a negative, temperature dependant coefficient denoting the strength of this interaction, $\tilde{d} \equiv d/l=1$ is the rescaled dipole length of the ion-pair, and the product $\delta \tilde{l}_B$ remains the Coulomb strength.

To ensure stability of the chain, $w_1 < 0$ demands the incorporation of a repulsive (positive) three body interaction term in the free energy\cite{muthu2004,arindam2010} of strength $w_3(>0)$, of the form, 
\begin{equation}
\frac{F_{55}}{k_{B} T}=\frac{2 w_3}{\tilde{l}_{1}^3}+\frac{w_3}{\tilde{l}_{13}^3}.
\label{F55tot}
\end{equation}
Once these two terms are added to the free energy in Eqs. \ref{free1} to \ref{freentotal}, for specific sets of parameters $w_1, w_3$, indeed collapse of the complex thus formed to sub-Gaussian sizes, is observed. Note that once the $F_{51}$ term in Eq. \ref{free5} can be ignored for sub-Gaussian sizes [$\tilde{l}_{13}$ (or $\tilde{l}_1$) $<1$], Eqs. \ref{F54tot} and \ref{F55tot} imply that $\tilde{l}_{13}$ (or $\tilde{l}_1$) scales as $N^{-1/3}$, leading to $R_{g12}^2=N l^2 \tilde{l}_{13}/6$ (Eq. \ref{Rgs}) going as $N^{2/3}$ or $R_{g12}$ as $N^{1/3}$\cite{arindam2010}. For example, with $w_1=-0.0057, w_3=3.3$, the expansion factor for the complex, $\tilde{l}_{13}$, is 0.96 and 0.52, respectively, and the free energy of complexation due to dipolar interactions, $\Delta (F_{54}+F_{55})/k_B T$, is -5.5 $k_B T$ and 21 $k_B T$, respectively, for $\tilde{l}_B=3$ and 5 ($\delta=3.0$). This energy is negligible compared to that of the free ion entropy ($F_2$) and ion pair enthalpy ($F_4$), also observed in simulations\cite{liu2002,winkler1998}. The individual polyions take sub-Gaussian and more compact forms for high, and the full complexes for moderate to high, Coulomb strengths. The isolated chains become sub-Gaussian due to dipolar correlations for $\tilde{l}_B > 4.28$, but still may not take a fully compact spherical shape until at very high Coulomb strengths\cite{liu2002,zhaoyang2006,winkler1998}, and its solubility and dynamics can only be predicted qualitatively. For example, coiled polyions are observed to stretch before overlap\cite{dzubiella2016} or globuler clusters split\cite{chen2022} (although the dynamics becomes much slower at a lower charge asymmetry).  $w_1, w_3$ remain a bit arbitrary, but even ignoring their temperature dependence and higher order virial terms, we did not get values of dipolar attractive energy competitive to $F_2$ and $F_4$. In principle, the parametric space of $w_1, w_3, \tilde{l}_B, N$, and $\delta$ can be thoroughly explored for which the full complex takes sub-Gaussian globular form for a larger range of Coulomb strengths keeping the isolated chains extended, as is observed in simulations. $w_1, w_3$ may be obtained from microscopic calculations including their temperature dependence. We must note, however, that dipolar interaction is just one way to address the charge correlations, and it ignores the part of cooperative interactions due to other topological conformations of the individual chains allowed by their interpentrability and intrinsic flexibility, such as in a 'scrambled egg' model. Still, it is the energies related to only the number of ion-pairs ($F_4$) and released counterions ($F_2$) which matter the most for a substantially large parameter range, and which is of the same order for the ladder and scrambled models with interpenetrating chains\cite{semenov2012}.

\subsection{Effect of Added salt}

\begin{figure}[h]
\centering
\includegraphics[height=3.8cm,width=4.43cm]{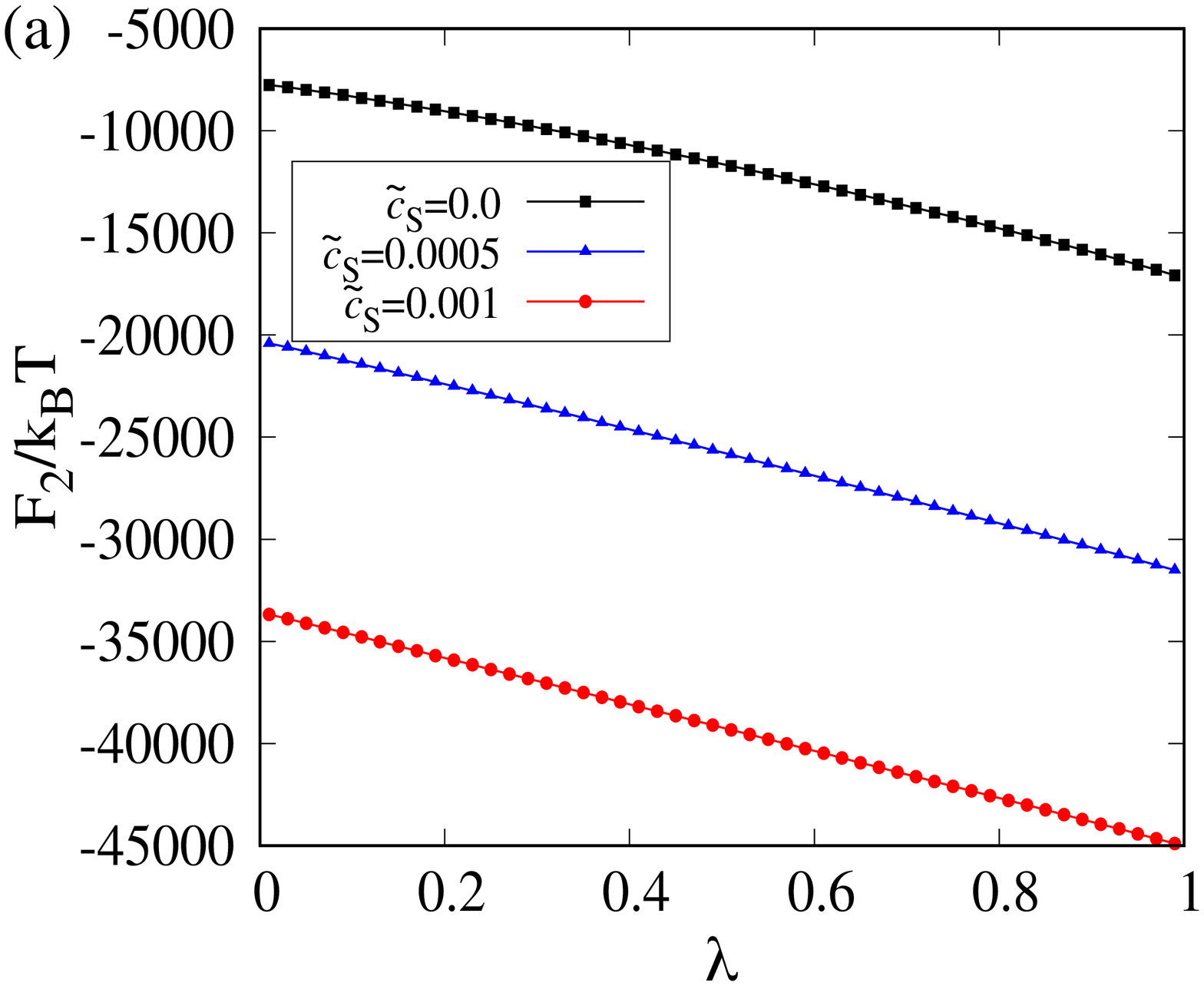}~~
\includegraphics[height=3.8cm,width=4.43cm]{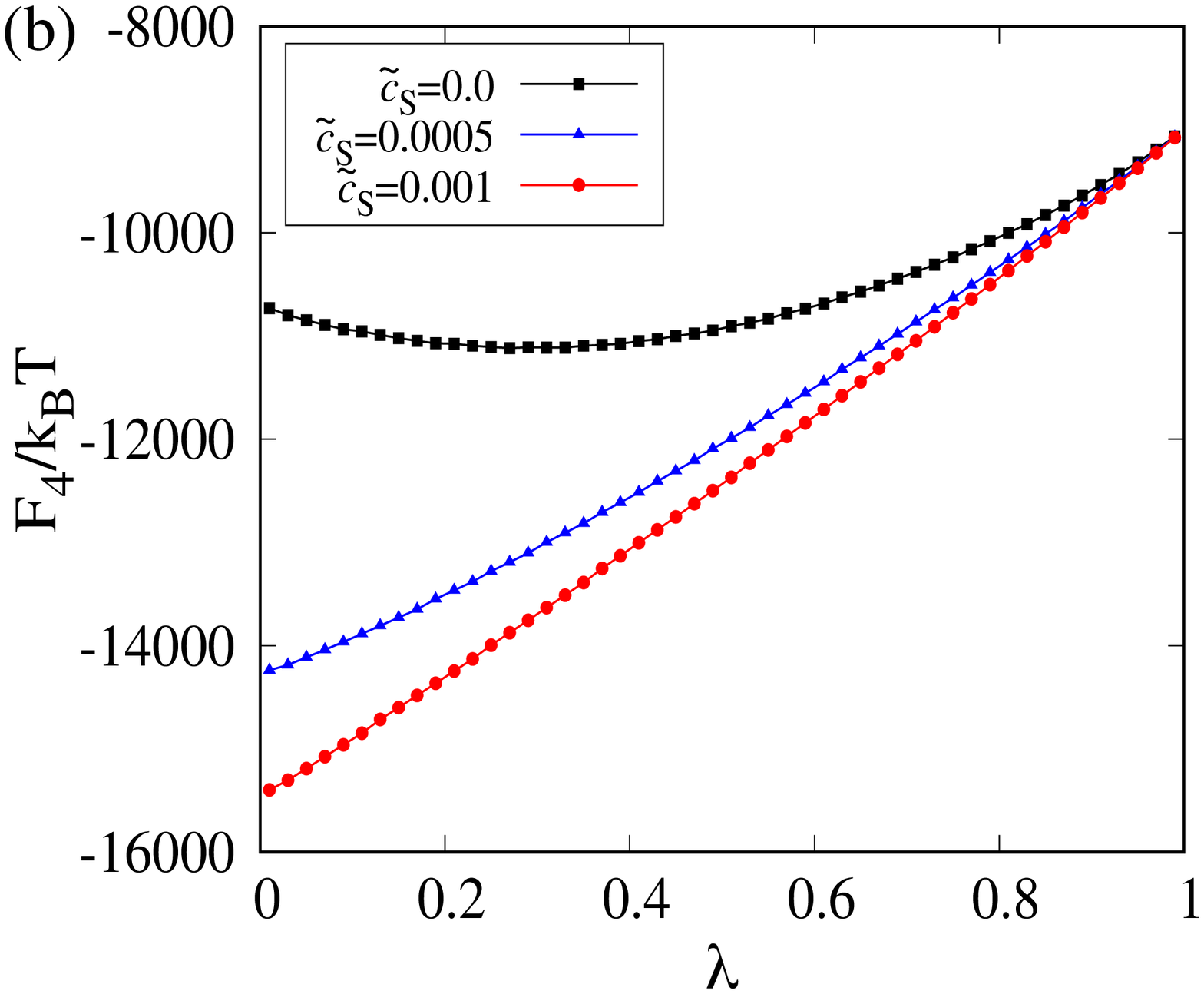}\\
\caption{Comparison of (a)free counterion entropy ($F_2/k_B T$), (b) bound ion pair enthalpy($F_4/k_B T$) for three different salt concentrations, $\tilde{c}_s=0,0.0005,0.001$. Parameters taken were: $\tilde{l}_B=3.0$, $\delta=3.0, N=1000$.}
  \label{salt-compare}
\end{figure}

Salt reduces the effect of all types of electrostatic interactions including the entropy gain from
released counterions\cite{muthu2002,muthu2004,lee2009,arindam2010,tirrell2013,mazumder2014,mitra2017,chen2022} in PE complexation. However, salt effects can be more subtle. One needs to note that it is not
the absolute values of enthalpy ($F_4$) or entropy ($F_2$), rather their differences between
the separated and complexed states, which drive the process. Although the total
entropy of free ions in both states increase with salt, the difference seems to remain similar in our model for low salts (to which this model is restricted to, as discussed later). However,
for enthalpy, with higher salt the separated chains will start with more counterions condensed on them (hence, a lower free energy with around $2N$ ion-pairs). Once they complex, almost half the ion-pairs will be lost (around $N$ ion-pairs). As a consequence, the enthalpy will increase (absolute value will decrease). Therefore, the enthalpy of complexation will increase with salt, eventually saturating, and would be progressively prohibitive to complexation. We observe these trends in Fig. \ref{salt-compare}.

We need to put a word of caution here regarding the theory we are using for salt.
It is well-known that the Debye-H\"uckel (DH) theory in its simplest form ($F_3$, Eq. \ref{free3}) 
would be strictly valid only for the regime the Debye length, $\kappa^{-1} \geq \tilde{l}_B$, the Bjerrum length. 
Therefore, results obtained with the theory of this work for 
high salts will be questionable. Ideally, from the expression of $\kappa$ we note that the
limit for the salt $\tilde{c}_s \leq (1/8 \pi \tilde{l}_B^3)$ must hold for the DH theory to be valid. For the highest value of 
$\tilde{l}_B$ (=5) that we have chosen, $\tilde{c}_s=0.000318$. 
The limits of salt, $\tilde{c}_s$, in our numerical calculation has been set to 
$\tilde{c}_s=0.001$, which is still beyond the range of DH-theory.

Using $\tilde{c}_{s} = 0.6 l^3$ $x$ molar, where $l$ is in nm, and using the Kuhn length $l=0.25~\mathrm{nm}$, such as it is for NaPSS, we get $x=0.106$, or $106~\mathrm{mM}$ of salt  corresponding to $\tilde{c}_s=0.001$, the highest value of salt that we have used. This DH limit of salt will vary for different PE pairs. We note that typical simulations\cite{zhaoyang2006} use a higher maximum salt (500 mM), which may not be enough to dissolve the complex, seen recently\cite{chen2022}. Diminishing entropy gains at higher salt concentrations seem to be the main reason for such dissolution.  However, as discussed above, one needs an approach more sophisticated than Debye-H\"uckel theory to deal with higher salts, which is beyond the scope of this current work.

\subsection{Entropy and Enthalpy of Complexation}

\begin{figure}[h]
\centering
\includegraphics[height=3.5cm,width=4.5cm]{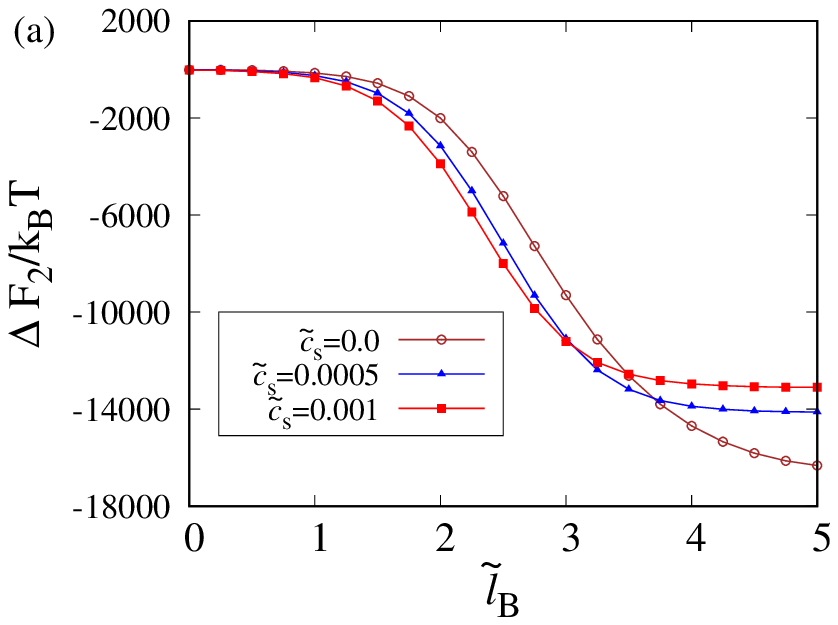}~~
\includegraphics[height=3.5cm,width=4.5cm]{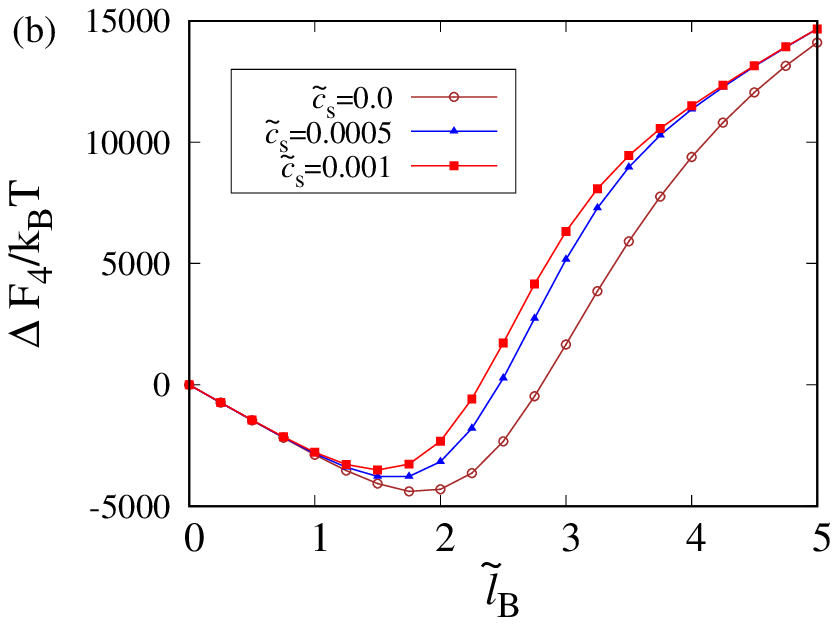}\\
\includegraphics[height=3.5cm,width=4.5cm]{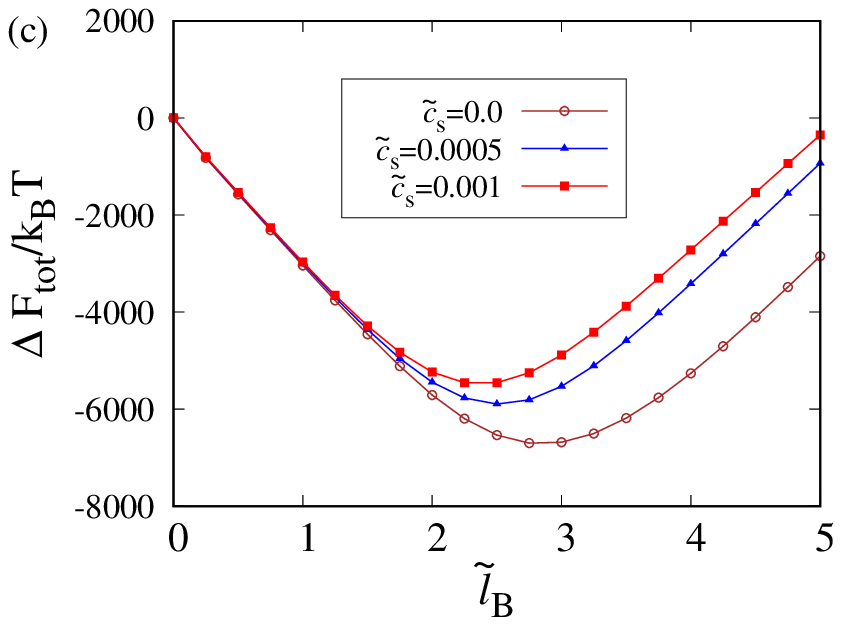} 
\caption{Variation of (a) free energy of complexation of counterion release ($\Delta F_2/k_B T$), (b) enthalpic free energy of complexation of Coulomb ion-pairs ($\Delta F_4/k_B T$) and (c) total free energy of complexation ($\Delta F_{\mbox{tot}}/k_B T$) with increasing $\tilde{l}_B$, for three different salt concentrations $\tilde{c}_s=0,0.0005,0.001$. Parameters taken were: $\delta=3.0$ and $N=1000$.}
  \label{entropy-enthalpy}
\end{figure}

We have already established, for a few discrete values of the Coulomb strength 
($\delta=3, \tilde{l}_B = 1$ to 5), that free ion entropy ($F_2$) and bound-pair energy ($F_4$) 
dominate the equilibrium free energy (see Fig. \ref{free1free2}, \ref{free3free4}) for all values of overlap $\lambda$. Further, $F_2$ is always supportive to complexation, more
so for higher Coulomb strengths, but enthalpy may be supportive (at low Coulomb strengths) or 
prohibitive (at high Coulomb strengths). In this section we briefly present in 
Fig. \ref{entropy-enthalpy} the variation of the free energy of complexation (that is the difference between $\lambda\equiv n/N=0$ and 
$\lambda=1$ states) due to the counterion release entropy ($\Delta F_2/k_B T$), Coulomb ion-pair enthalpy ($\Delta F_4/k_B T$), and the total free energy of complexation ($\Delta F_{\mbox{tot}}/k_B T$), with the Coulomb strength 
(for the same range, $\delta=3, \tilde{l}_B = 1$ to 5), and for three different salt values,
$\tilde{c}_s=$ 0, 0.0005, 0.001. $F_{\mbox{tot}}$ is given by Eq. \ref{freentotal} and $F_i$ ($i=$1,6) are given through Eqs. \ref{free1} to \ref{free6}. We choose $F_2$ and $F_4$ specifically, compared to the simulations\cite{zhaoyang2006} which include all contributions respectively, to emphasize that these are the major ones which account for the total free energy of complexation. 

We noted above that for such a low salt concentration we can not effectively differentiate the entropy gain of counterion release for different salts [Fig. \ref{entropy-enthalpy}(a)]. However, the gain 
monotonically increases with the Coulomb strength as expected\cite{zhaoyang2006,whitmer2018macro}. 
The enthalpy of complexation due to Coulomb ion-pairs shows a better variation. 
Once salt ions are present, both separated PE chains have most of their 
counterions condensed before complexation\cite{muthu2004,zhaoyang2006,lee2009,arindam2010,rumyantsev2018,zhen-gang2018,panyukov2018}. Therefore, the enthalpy gains progressively more negative values (which we call enthalpy 'gain') with more number of ion-pairs (monomer and counterion) to begin with at higher salt. After complexation, all such monomer-counterion pairs are dissociated and oppositely charged monomer-monomer pairs form. Therefore, both for high $\tilde{l}_B$ at low salt and high salt at moderate $\tilde{l}_B$, almost half of the ion-pairs are reduced after complexation, and the enthalpy of complexation opposes the process
substantially. In other words, the monomer-monomer pairs are much less in number compared to the monomer-counterion pairs for high salt or high Coulomb strength. This is a positive change for enthalpy of complexation (more negative before complexation to less negative after complexation), which we call the enthalpy 'loss' due to complexation. This will correspond to positive values of enthalpic free energy of complexation, $\Delta F_4/k_B T$ [for $\tilde{l}_B \ge$ around 2.3, depending on the salt, in Fig \ref{entropy-enthalpy}(b)].   Therefore, the free energy drive of complexation progressively decreases for higher salt [Fig. \ref{entropy-enthalpy}(c)] - due to positive $\Delta F_4$ for moderate Coulomb strengths and positive $\Delta F_2$ for high Coulomb strengths (around $\tilde{l}_B=5.0$).  Finally, the drive vanishes for a sufficiently high Coulomb strength, 
that has been profoundly observed in all PE complexation literature\cite{dautzenberg2002,zhaoyang2006,spruijt2010macro,rumyantsev2018,zhen-gang2018,panyukov2018,ong2019,shakya2020,chen2022}. We, however, have not covered in Fig. \ref{entropy-enthalpy}(c) the high $\tilde{l}_B$ range for which a loss in free energy is visible (positive values of $\Delta F_{\mbox{tot}}/k_B T$). In that limit the salt concentration we have used becomes questionable within the Debye-H\"uckel treatment of ionic fluctuations. Regardless of this limitation, the Debye-H\"uckel theory still remains a popular way of treating ionic fluctuations \cite{qin2014,Jha2014,pablo2016,radhakrishna2017,adhikari2018}, and is used to find the basic results in our theory.

The threshold $\tilde{l}_B$, say $\tilde{l}_B^\star$, that identifies the boundary between enthalpy- and entropy-driven complexation can be estimated by equating $\Delta F_2$ and $\Delta F_4$, where $F_2$ and $F_4$ are given by Eqs. \ref{free2} and \ref{free4}, respectively. We get a transcendental equation,
\begin{equation}
\tilde{l}_B^\star=\frac{2\left[\left(N+n_s\right) \log (\tilde{\rho}+\tilde{c}_s) -
\left(Nf+n_s\right) \log (f\tilde{\rho}+\tilde{c}_s)- N (1-f)\right]}{N \delta (1-2f)},
\label{lbstar}
\end{equation}
where, in this case, $\tilde{\rho}=N/(\Omega/l^3)$, and $f$ is given by $f^\star = f(\tilde{l}_B^\star)$, using Eq. \ref{chargeanalyticaln} with $n=0$. This expression is valid generally for all values of 
$\delta$ and $\tilde{c}_s$ explored in our work.

\subsection{Phase Diagram}

With increasing Bjerrum length, $\tilde{l}_B$, the entropy gain of complexation saturates (from no free ions when chains are separated to maximum number of free ions when fully complexed), but the enthalpy loss continues to increase
(linearly with $\tilde{l}_B$) [Fig. \ref{entropy-enthalpy}]. (It is actually the enthalpy loss that saturates, but the entropy gain becomes 
progressively insignificant with decreasing temperature. The comparative strengths of the above
effects apparently seem to be reversed, because the free energy is written in the units of $k_B T$). Threrefore, 
one may predict that for a certain Coulomb strength [i.e., a certain set of $\tilde{l}_B, 
\delta$ (for example, at a certain low temperature in a specific solvent for a specific pair of polyions)] 
the enthalpy loss shall offset the 
entropy gain, and complexation will be disfavoured. The monotonicity of the
downward trend of the total free energy versus overlap curve (change in $F_{\mbox{tot}}$ 
against $\lambda$) will get affected, and eventually it shall go upward, finally crossing the 
zero free energy of complexation, at which point complexation becomes unfavourable
compared to stable, isolated chains. 

Further, near the transition values of $\tilde{l}_B, \delta$, for a narrow
range of the parameter values, the free energy is expected to be minimum
for neither $\lambda=0$ (separate chains) nor $\lambda=1$ (complexed chains),
but for some intermediate value of $\lambda$ (partial overlap). In that range
only a fraction of oppositely charged monomers from both chains will form ion-pairs at true equilibrium.
The remaining monomers will remain in the dangling chain ends, again with a fraction of them having
their counterions condensed just as in a single, isolated chain. 

\begin{figure}[h]
\centering
\includegraphics[height=3.8cm,width=4.43cm]{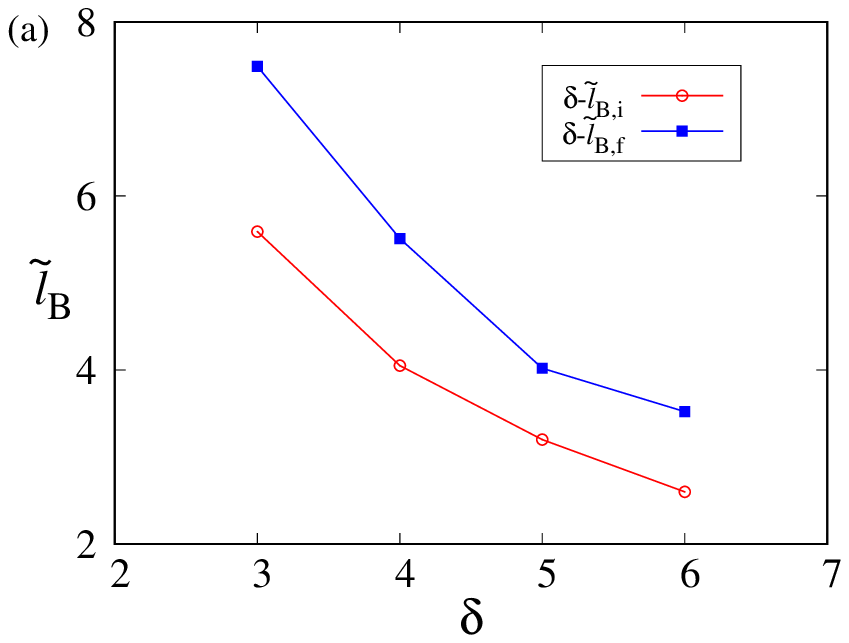}~~
\includegraphics[height=3.8cm,width=4.43cm]{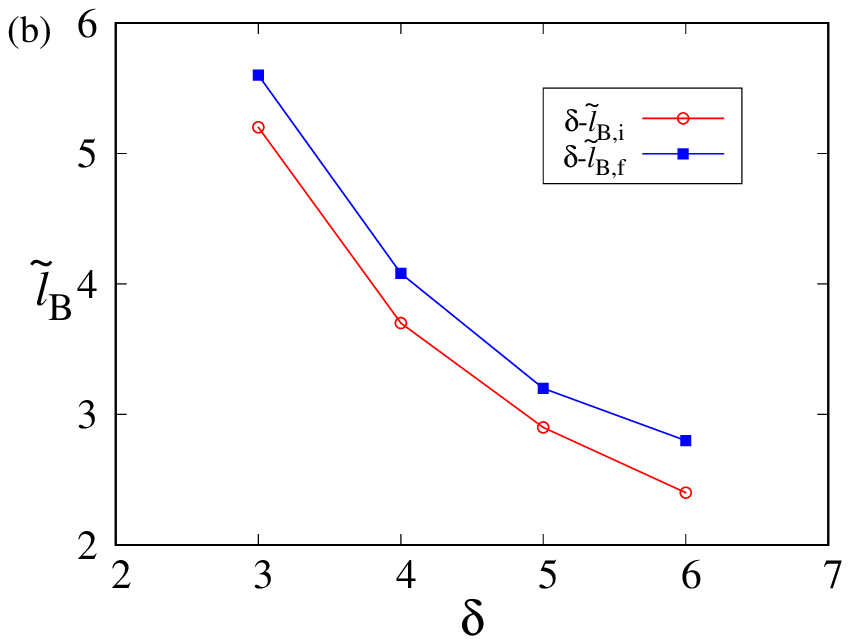}\\
\includegraphics[height=3.8cm,width=4.43cm]{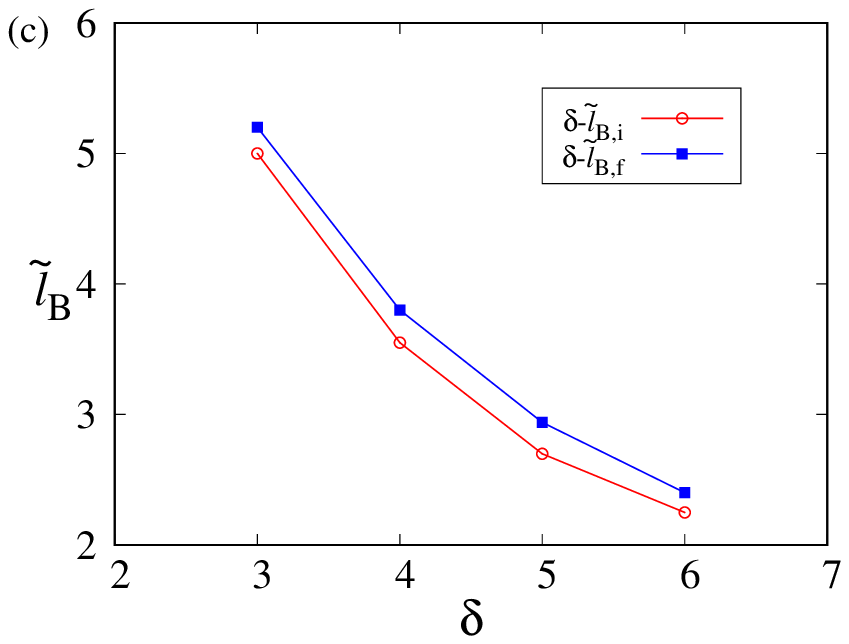}~~
\includegraphics[height=3.8cm,width=4.73cm]{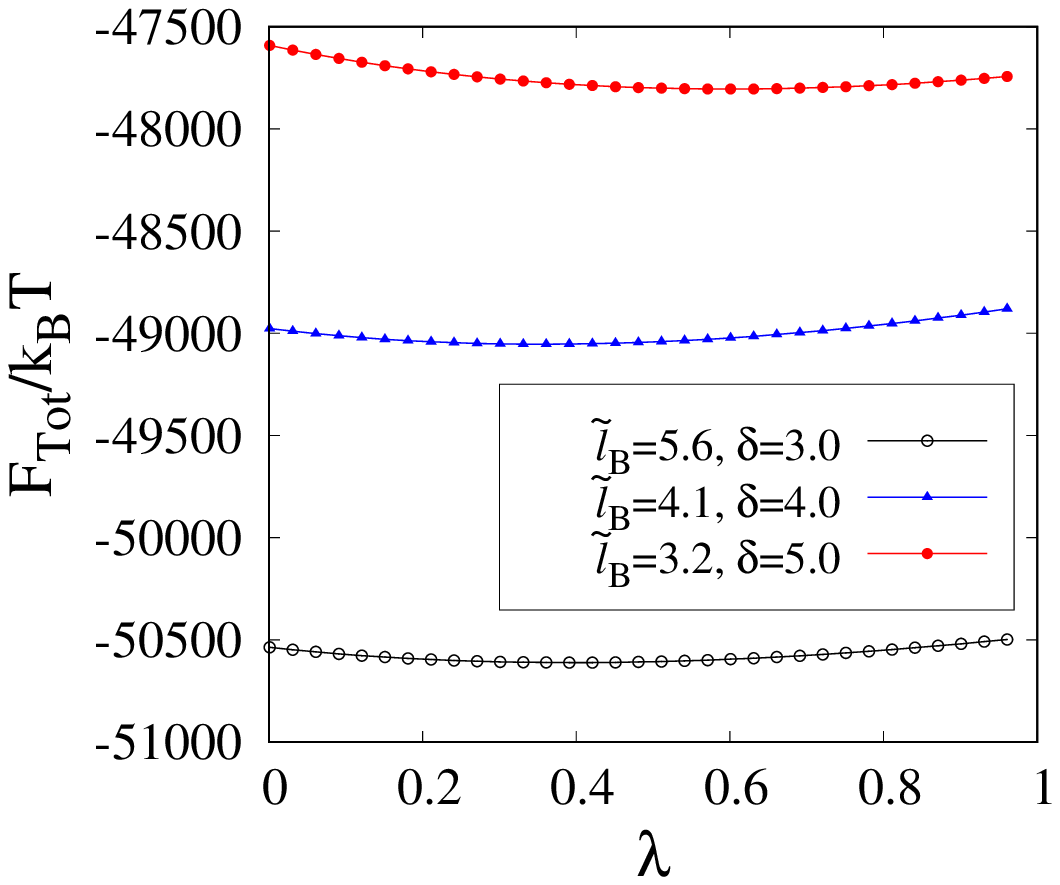}
\caption{$\delta-\tilde{l}_B$ phase diagrams showing the transition $\tilde{l}_B$'s, marking the partially complexed regimes for (a) $\tilde{c}_s=0.0$, (b) $\tilde{c}_s=0.0005$ and (c) $\tilde{c}_s=0.001$. To the left of
the red line (circles) the chains are fully complexed, whereas to the right of the blue line (squares) they are fully separated.The total free energy is plotted as a function of overlap $\lambda$, for three pairs of $\delta, \tilde{l}_B$ values. (d) shows existence of a minimum at intermediate values of overlap (between $\lambda=0$ and $\lambda=1$), for a fixed salt $\tilde{c}_s=0.0005$. Parameter values of $\delta$ and $\tilde{l}_B$ are taken from within the narrow band corresponding to partially complexed chains in Table \ref{phasediagtable}.}
  \label{phasediagram}
\end{figure}

With this prediction in mind we plot the transition temperatures 
$T_c$ (corresponding to $\tilde{l}_{B,i}$ in Table \ref{phasediagtable}), 
at which the complete complexation gives way to partial complexation
as the stable state, and $T_n$ (corresponding to $\tilde{l}_{B,f}$), 
at which the partial complexation gives way to fully separate chains as the stable state, for fixed values of $\delta$, for the same three salt concentrations
($\tilde{c}_s=0, 0.0005, 0.001$). Fig. \ref{phasediagram} 
shows results expected along the line of the discussion above. The $\delta-\tilde{l}_B$ parameter space does not include values of $\tilde{l}_B>8.0$ to avoid the possibility of small ion-pairing (see discussion in 'The full numerical results').

We must repeat the caution exercised in the discussion of salt effects. The DH theory
we have used is limited to low salts at high $\tilde{l}_{B}$. Hence, for a fixed salt,
the results at high $\tilde{l}_{B}$ would be questionable, and should only be accepted
qualitatively. One may, however, note that (Table \ref{phasediagtable}) for
the transition temperatures the value of the Coulomb strength 
($\tilde{l}_B \delta$) for different pairs of $\tilde{l}_B$ and $\delta$ stays constant 
to an excellent degree. This confirms
the significant and decisive role of electrostatics, in both controlling the enthalpy of 
bound ion-pairs and concomitant entropy of free ions both of which dominate to set up the equilibrium for
the system.    

It is worthy of note here that the Coulomb strengths applicable to the phase boundaries may be reasonably higher than modest experimental values applicable to known homopolymers. As mentioned before, the modest values of $\tilde{l}_B$ and $\delta$ can be obtained, for example, for NaPSS in water at room temperature, in which case $\tilde{l}_B \simeq 3$ and $\delta \simeq 3.5$. However, it may be possible to have other values of $\tilde{l}_B$ and $\delta$ in other types of systems [intrinsically disordered proteins (IDP)s, other biological polymers etc.] which can be addressed by these phase diagrams.
  
\begin{table}
\centering
\resizebox{\columnwidth}{!}{%
\begin{tabular}{|c|c|c|c|c|c|c|c|c|c|c|c|c|}
\hline & \multicolumn{4}{|c|}{$\tilde{c}_{S}=0.0$} & \multicolumn{4}{|c|}{$\tilde{c}_{S}=0.0005$} & \multicolumn{4}{|c|}{$\tilde{c}_{S}=0.001$} \\
\hline $\boldsymbol{\delta}$ & $\tilde{l}_{B, i}$ & $\tilde{l}_{B, f}$ & $\boldsymbol{\delta} \tilde{l}_{B, i}$ & $\boldsymbol{\delta} \tilde{l}_{B, f}$ & $\tilde{l}_{B, i}$ & $\tilde{l}_{B, f}$ & $\boldsymbol{\delta} \tilde{l}_{B, i}$ & $\boldsymbol{\delta} \tilde{l}_{B, f}$ & $\tilde{l}_{B, i}$ & $\tilde{l}_{B, f}$ & $\boldsymbol{\delta} \tilde{l}_{B, i}$ & $\boldsymbol{\delta} \tilde{l}_{B, f}$ \\
\hline & & & & & & & & & & & & \\
\hline $3.0$ & $5.6$ & $7.5$ & $16.8$ & $22.5$ & $5.2$ & $5.6$ & $15.6$ & $16.8$ & $5.0$ & $5.2$ & $15.0$ & $15.6$ \\
\hline $4.0$ & $4.1$ & $5.5$ & $16.4$ & $22.0$ & $3.7$ & $4.1$ & $14.8$ & $16.4$ & $3.6$ & $3.8$ & $14.4$ & $15.2$ \\
\hline $5.0$ & $3.2$ & $4.2$ & $16.0$ & $21.0$ & $2.9$ & $3.2$ & $14.5$ & $16.0$ & $2.7$ & $2.9$ & $13.5$ & $14.5$ \\
\hline $6.0$ & $2.6$ & $3.5$ & $15.6$ & $21.0$ & $2.4$ & $2.8$ & $14.4$ & $16.8$ & $2.3$ & $2.4$ & $13.8$ & $14.4$ \\
\hline & & & & & & & & & & & & \\
\hline
\end{tabular}%
}
\caption{Tabulation of the transition temperatures($\tilde{l}_B$s) and the product of $\delta-\tilde{l}_B$ for three different salt concentrations $\tilde{c}_s=0,0.0005~\mathrm{and}~0.001$ for $N=1000$. 
$\tilde{l}_{B,i}, \tilde{l}_{B,f}$, respectively, correspond to fully-complexed to partially-complexed and
partially-complexed to fully-separated phase boundaries.}  
\label{phasediagtable}
\end{table}


To elucidate the partially complexed state, we plot the free energy against $\lambda$ for a few pairs of $\delta, \tilde{l}_B$ values [Fig. \ref{phasediagram}(d)] applicable for such states. For most of the parameter space, it is found that the lowest free energy state is either fully separate chains ($\lambda=0$), for high Coulomb strengths, or fully complexed chains ($\lambda=1$), for low Coulomb strengths. There we find a very narrow range of parameter space, however, for which the gain in $F_2$ and loss in $F_4$ would just balance out, and the gain in total free energy is maximum for (or, in other words, the lowest free energy state is at) intermediate values of $\lambda$ [Fig. \ref{phasediagram}(d)]. This will give rise to this narrow band in parameter space within which the partially complexed chains are the stable ones. Both in experiments and in simulations it may be of interest to explore pairs of PEs which will show partially complexed states.

The excluded volume parameter $w$, the negative value of which tend to minimize the surface energy in collapsed, sub-Gaussian polymers, is taken to be zero in our work, implying that surface tension is neglected. However, the $w_{12}$ term in $F_{52}$ (Eq. \ref{free5}) is similar to the $w_1$ term in $F_{54}$ (Eq. \ref{F54tot}), used for dipolar correlations. The energetic effect of such terms ($w_{12}<0$), compared to the electrostatic contributions $F_2$ and $F_4$, would be negligible (see section `Collapse of Chains in Complex'). Indeed the phase boundaries for the partially complexed states, no matter how close in terms of Coulomb strength, are found to survive the set of modest values of $w_1, w_3$ we have used in Eqs. \ref{F54tot} and \ref{F55tot}. With these dipolar correlations, or alternatively, surface tension contributions ($w_{12}<0$), both the dangling chains and the complex are in sub-Gaussian conformation, for which we can only predict the thermodynamics in this work, but are unable to ascertain whether the phases will be kinetically achievable.

\subsection{Comparison to Simulations}

In the discussion of Fig. \ref{entropy-enthalpy} we pointed out the good qualitative agreement of (a) entropy, (b) enthalpy and (c) free energy of complexation between our theory and the simulations published in the literature. To discuss in more details, we note that the enthalpy of complexation as a function of the Coulomb strength has a similar trend between the simulation (Figs. 8 and 9 in 
Ref. 17), which also considers fully ionizable PEs, and our theory [Fig. \ref{entropy-enthalpy}(b)]. For quantitative matching we first take the salt-free results. One can identify the Coulomb strength at which the enthalpy crosses value zero from negative and becomes increasing positive. In the simulation it is $\Gamma \equiv l_B/l_0 \simeq 2.5$, where $l_0$ is the bond length, whereas in our theory it is $\tilde{l}_B \equiv l_B/l \simeq 2.7$, where $l$ is the Kuhn length. For a fully ionizable flexible chain, $l_0=l$, implying an excellent match for the simulation and our theory, although $\delta$, equal to 3 in theory, remains uncertain in simulations. The maximum gain in the enthalpy (for $\Gamma, \tilde{l}_B$ between 1 to 2 in both simulation and theory) is around $300 ~ k_B T$, i.e., $\simeq 5 ~ k_B T$ per bound pair (as $N=60$) in simulation. In comparison, it is around $5000 ~ k_B T$, i.e., $\simeq 5 ~ k_B T$ per bound pair ($N=1000$) in our theory. The enthalpy gain reduces with salt in both simulation and theory for low to moderate $\Gamma,\tilde{l}_B$. However, the scales for salt concentration are different between theory and simulation, and high salt results are only qualitatively correct for both, less for the theory. For a high Coulomb strength ($\tilde{l}_B=4$), the Coulomb energy loss from simulation is around $400~k_BT$, giving $\simeq 7~k_B T$ per bound pair, while from theory it is around $9000~k_BT$, leading to $\simeq 9~k_B T$ per bound pair. Here, though the enthalpy per bound pair is still close from both the approaches, an exact one-to-one matching may not be appropriate at all Coulomb strengths since it is only the bound ion-pair enthalpy for the theory (which forms an overwhelmingly significant part of enthalpy) but all Coulomb energies for the simulation. Similar to our theory, Ref. 17 too finds that no complexation occurs for high $\Gamma$ ($\tilde{l}_B$), although it could not be assigned a reason due to long simulation times.

For the entropy of complexation, shown in Fig. 11 (Ref. 17) and Fig. \ref{entropy-enthalpy}(a) here, we note that the entropy is calculated in Ref. 17 from theoretical considerations (including mean field models) taking input from simulation data. The trend of entropy gain, however, shows a similar feature for both simulation and our theory [the signature of entropy is opposite between Ref. 17 and our theory, because Ref. 17 plots the absolute value of entropy $S$ (in terms of $k_B$), while we plot $-TS$, the free energy contribution from it (in terms of $k_B T$)]. Quantitatively, the maximum entropy gain (for $\Gamma, \tilde{l}_B=$ 4.5-5.0) from the simulation (salt-free) is approximately $950~k_BT$ which gives $\simeq 16~k_B T$ gain per bound pair, while in our case [Fig. \ref{entropy-enthalpy}(a)] we get the maximum gain of about $16000~k_B T$ resulting in a similar gain of $16~k_B T$ per bound pair. 

The total free energy of complexation [Fig. \ref{entropy-enthalpy}(c) in our theory and Figs. 8 and 12 in simulation, Ref. 17] has a similar match. For the salt-free curve, the maximum free energy gain (occuring in both simulation and theory for $\Gamma, \tilde{l}_B$ between 2.5 to 2.7) is noted to be $\simeq 650~k_B T$, i.e., $\simeq 11~k_B T$ per bound pair ($N=60$) from simulation. From our theory it is noted to be $\simeq 7000~k_B T$, giving approximately a $7~k_B T$ free energy per bound pair ($N=1000$). This quantitative difference is due to a lower value of entropy gain in theory ($6~k_B T$ per bound pair) compared to simulations ($10~k_B T$ per bound pair) at $\Gamma, \tilde{l}_B$ around 2.5 to 2.7. The trend of free energy gain, rising with $\Gamma, \tilde{l}_B$ up to values 2 to 3 and then declining is very similar between simulation and our theory. We may note here that the phrases `potential of mean force' and `free energy' have been interchangeably used in the simulation papers.

In Ref. 80, PE complexation of two oppositely charged polyions in nominal salt (20 mM) is simulated using a similar potential of mean force (PMF) method. The simulation finds similar sliding approach of PEs for which the monomers form bound ion-pairs sequentially. It uses fully ionizable monomers, although the total charge has been distributed among the monomers apparently in a mean field way. Dielectric constant $\epsilon=78$ and ion size=0.3 nm are very similar parameters between the simulation and our theory, making them better comparable. We must note that the results from our theory are for salt-free case. The equivalent dimensionless Manning parameter in this simulation is $\xi=l_B(\lambda_c/e)$, where $\lambda_c \simeq e/l_0$, leading to $\xi=l_B/l_0$ (same as $\tilde{l}_B$ in our theory and $\Gamma$ in simulation of Ref. 80). The gain in entropic free energy, calculated from their preliminary theory, is around 4.7 $k_B T$ per released ion ($\xi=2.31$). In our theory, for $\tilde{l}_B=3$ it is 8000 $k_B T$, for which around 1200 counterions are released in complexation (as the degree of ionization is 0.4 for the $N=1000$ chain before complexation [Fig. \ref{chargesizeanal}(a)], same as that in another simulation\cite{peng2015} similar to Ref. 17 for which $N=60$ and $\Gamma=2.8$), i.e., around 6.7 $k_B T$ per released ion. This is of order the simulation result but expectedly higher, as $\tilde{l}_B$ is higher for the theory. Most importantly, the total free energy (or the PMF), from the onset of complexation, is linearly decreasing with centre-of-mass (COM) distance between the chains (equivalent to our overlap parameter $\lambda$) for both theory (Fig. \ref{free6free_tot}b) and simulations (Refs. 80 and 117), which implies a constant force of attraction during complexation. The linear PMF (constant force of complexation) is found to be not valid for higher $\Gamma$ (Ref. 117), the same trend being observed in our theory too where it becomes slighly concave upwards [Fig. \ref{freetanalfig} or \ref{free6free_tot}(b)]. The free energy gain increases with $\xi$ in simulation of Ref. 80 (and with $\tilde{l}_B$ in our theory) up to around $\xi=2.31$, beyond which the simulation results are not available. It is dominated by enthalpy for $\xi<1$ and counterion-release entropy for $\xi>1$ (similar to our low and high $\tilde{l}_B$ results in theory). The total free energy of complexation is around $155~k_B T$ at a Coulomb strength of $\xi=2.31$ in simulation (Ref. 80), which implies $\simeq 6~k_B T$ per bound pair ($N=25$). In our theory, for a slightly higher Coulomb strength [$\tilde{l}_B=3$, Fig. \ref{entropy-enthalpy}(c)]  the value is 7000 $k_BT$ leading to $\simeq 7~k_B T$ gain per bound pair.

For increasing salt, more condensation in isolated chains leads to more ion-pairs lost in complexation. Consequently, enthalpy gain decreases and eventually changes to enthalpy loss at a lower $\Gamma$ ($\tilde{l}_B$) for both simulations (Fig. 9, 10 in Ref. 17) and the theory (Fig. \ref{entropy-enthalpy}b). Further, the Coulomb energy is slightly positive for separated chains at low $\Gamma$ ($\tilde{l}_B$) ($ \sim 1.0$, Fig. 6, Ref. 17 and Fig. \ref{free51free53}b in theory), at which the condensation is negligible, but uncompensated monomers cost a repulsive electrostatic energy ($F_{53}$, Eq. \ref{free5}). The free energy of interaction of free ions is accounted for by pure, pairwise Coulomb energy in simulations and charge density fluctuations (Debye-H\"uckel) in the theory ($F_3$, Eq. \ref{free3}). Both, for the salt-range explored, do not show much change in complexation (increases up to only 1-2 $k_B T$ per bound-pair for even the highest Coulomb strengths and salts we have explored in the theory, Fig. \ref{free3free4}a). Accounting for higher salts will require a more sophisticated theory, as discussed before. Although weaker pairing of monomers is observed for low $\Gamma$ (the radial distribution function, Fig. 4a, Ref. 17) - at $\Gamma=2.0$ there are around 70\% pairs with the assumed cut-off (the rest of the pairs not completely separated due to connectivity) - a reasonable match with the theory of complete binding, and also a significantly broad counterion distribution\cite{chen2022}, support the two-state model of ours. 

The dielectric mismatch, $\delta$, however, enhances the enthalpy loss of Coulomb pairs in the theory (15 $k_B T$ per monomer pair, Fig. \ref{entropy-enthalpy}b) compared to simulations, which include all contributions (8 $k_B T$, Fig. 9, Ref. 17) at higher Coulomb strengths ($\Gamma=\tilde{l}_B=5.0$). We note that the DH free energy of complexation ($\Delta F_3$) gains about 2 $k_B T$ at high $\tilde{l}_B$ that explains the discrepancy to some extent. The match is better (around 5 $k_B T$ gain) for low $\Gamma$ (see above). The enthalpy loss due to Coulomb pairs increases linearly with high Coulomb strengths which include the factor of $\delta$ for the theory. Further, in simulations\cite{zhaoyang2006}, monomers in the complex interact attractively, more with more salt and higher Coulomb strengths, with surrounding counterions of opposite charge with a density peak nearby\cite{chen2022}. This additional negative enthalpy decreases the loss, but is entirely absent in the two-state model. The final complex, however, is completely devoid of counterions for both. For low Coulomb strengths, more salt induces more counterion condensation in theory, leading to a decrease in enthalpy gain. This is explained in terms of screening in simulations\cite{zhaoyang2006}, but the latter using an implicit solvent method, and without a dielectric mismatch, underestimates the counterion condensation. An explicit atom simulation may shed light on this aspect.

The simulations\cite{zhaoyang2006} calculate the entropy of mixing of counterions and polymers, but, unlike the theory, which identifies the ideal gas volume entropy of free ions to be the most significant contribution, ignore the translational entropy of condensed ions. Two competing effects of increasing salts - that it increases condensation of counterions in isolated chains that are released to provide positive entropy of complexation, but it also reduces the entropy gain per released ion in complexation - may lead to a marginal change in the entropy of complexation at low $\Gamma$ (Fig. 11, Ref. 17, simulations) and $\tilde{l}_B$ (Fig. \ref{entropy-enthalpy}a, the theory). Either due to condensation of majority of counterions in isolated chains (in the theory for $\tilde{l}_B>3.5$) or weak variation in condensation (possibly due to the absence of dielectric mismatch) (in simulations for around $\Gamma=2.0$ and higher) the entropy of complexation decreases with salt due to the latter effect. Despite using a low salt (100 mM), the decrease in free energy gain at high $\tilde{l}_B$(=5) is more effected by the substantial loss in entropy (3 $k_B T$ per monomer pair) than in enthalpy (around 0.5 $k_B T$) (Fig. \ref{entropy-enthalpy}), similar to Ref. 17, Figs. 9 and 11. Both the theory (Fig. \ref{entropy-enthalpy}c) and simulations (Fig. 12, Ref. 17) predict that the highest salt to dissociate the complex will be required for medium Coulomb strengths ($\tilde{l}_B, \Gamma =2.0-3.0$), and  the maximum free energy gain shifts towards lower $\tilde{l}_B, \Gamma$ with higher salt. A more careful treatment of free ion charge correlations in the theory would allow better predictions for higher salts. 

Experiments\cite{vitorazi2014} too estimate the thermal energy per titrating charge in PE complexation, and find the enthalpy costs of the reactions to be of order $+1~k_B T$ (which is a bit low compared to simulations and our theory), whereas the gains in entropy are of order $+10~k_B T$ (close to simulations and our theory). It is, however, not clear whether the enthalpy arises from the pairing of the opposite charges only or other contributions are also significant. Changes in free energy are of order 1-10 $k_B T$ per titrating ligand (close to simulations and our theory).

Our theory suggests that simulations which would measure the change in entropy directly (as compared to mixed theoretical models) may give more insight. In addition, simulation thermodynamics at high Coulomb strengths (high $\Gamma = l_B/l_0$) and at different concentrations ranging from highly dilute to reasonably dense may be instructive, and may explore the partially complexed states as we do find in our theory.

\section{Conclusions}

Polyelectrolyte complexation not only has enormous application but also is theoretically challenging, due to a complex interplay of entropy and internal energy (enthalpy) of several components. In this work, we have proposed a minimal theory for the complexation of two symmetric, fully-ionizable (strong), and oppositely charged polyelectrolyte chains in a dilute solution in the presence of low salt. The two-chain PE complexation is widely considered as an important first step in the formation of complex coacervates. We may note that this simple two-chain assembly, fundamental to polymer complexation, is still a complex,
multicomponent system that deals with several energy and entropy contributions, effects of many of which are
intertwined. They include conformational free energy of polyions, translational entropy of free and condensed ions,
correlation free energy of density fluctuations of free ions, Coulomb energy gain of condensed ion-pairs of both types,
screened Coulomb interaction among charged monomers within and between chains, excluded volume effects, poor
solvent effects due to dipolar attractions etc. Further, it is not the absolute values of such quantities but their
changes during complexation that dictate the equilibrium.

It has been observed for most strong PE systems that the free ion entropy and the bound ion-pair energy dominate the equilibrium of the system. The aim of this work was to 
explore the same physics, quantify the relative importance of such effects in this two-chain system, and identify the thermodynamic driving (or opposing) forces of complexation. In the model, the intermediate state in complexation is made of the complexed middle part and dangling polyion chains on either side. The free energy of the problem is constructed following the Edwards Hamiltonian of a single PE within the uniform expansion model. The formalism is capable of capturing free energy contributions, in details, arising from the entropy of free ions, conformations of polymer chains, and electrostatic interactions among charged species, as well as from basic charge correlations due to dipolar attractions. The charge and size of the dangling chains and the free energy components are calculated, as functions of degree of overlap of chains within quasistatic approximation and added salt, for different Coulomb strengths.
  
The primary challenge of this minimal theory has been to identify approximations which are physically acceptable, as well as which simplify the approach and find major energetic contributions to PE complexation. The close match between the analytical and numerical results lends credence to the conjectures and appoximations. It is indeed observed that, for PE complexation as well, free ion entropy and bound ion-pair electrostatic energy dictate the equilibrium. With this understanding, the self-consistency in the free energy in terms of size and charge can be decoupled, which enabled us to identify a significant part of the free energy, ignoring the minor contributions, resulting in an analytical expression 
for the charge that directly determines the size of the dangling chains. This method eventually leads to an analytical formulation of the free energy of complexation, as well as of the enthalpy and entropy. It is found that the enthalpy of complexation of bound ion-pairs is supportive of complexation at low,
but opposing at high, Coulomb strengths. The entropy of complexation of free ions is always supportive, more so at higher Coulomb strengths, so much so that it can overcome the negative enthalpy, and drive the complexation handsomely for a substantial moderate range of electrostatic interactions. Eventually, complexation is energetically disfavoured at high electrostatic strengths, due to the prohibitive negative enthalpy. We calculate the threshold Coulomb strength that delineates the entropic and enthalpic gains as the major drive.

The drive for complexation as predicted by the theory, and the clear dominance of free ion entropy and bound-pair energy, apply excellently for the complexation in a two-chain system for a large range of dilutions, but will progressively deviate for denser solutions. For the latter, the free ion entropy will decrease due to a smaller available volume, and the translational entropy of the condensed small ions may rise considerably due to availability of multiple chains (this aspect is not explored in this work).   

To complete our analysis, we have reported full numerical results for the total free energy including all components, in addition to the two major thermodynamic drives. The analytical and full numerical results are found to be quantitatively close, as expected. We have
also observed that salt reduces the complexation drive for our model, in agreement to literature. Phase diagrams identifying three stable phases -
fully complexed, partially complexed, and fully separated chains - have been constructed as functions of different electrostatic strengths, and for different low salt concentrations. Small ion-pairing is found to be absent for the parameter space explored, because of the presence of dielectric mismatch. The complex collapses to sub-Gaussian sizes if charge correlations in the form of dipolar attractions are considered. However, the energies related to such conformational changes remain negligible.

The entropy, enthalpy, and free energy of complexation obtained as functions of the Bjerrum length, and the potential of mean force (free energy) as a function of the chain separation (overlap), are compared to known simulation results from the literature, and we find significant quantitative match. This is supportive of the mean-field model with basic charge correlations we propose. However, charge correlations and its effects on entropy and enthalpy at high salts need to be addressed by a theory more advanced than simple ion screening described by the Debye-H\"uckel theory. We suggest that simulations of PE complexation at high Coulomb strengths may be explored to identify the partially complexed states seen in our theory. 

We summarize that, the free energy used to model PE chains and solutions, and
to match experimental results,
is successfully used to look at a fundamental aspect of the problem and develop an analytical scheme to predict the thermodynamics of PE complexation. This uniform expansion model leads to a free energy made of additive entropic and enthalpic terms,
that in turn generate quantitative estimation of thermodynamic quantities, directly comparable to simulations on
strong polyelectrolytes. The free energy reasonably incorporates most contributions in the system, and, most
importantly, treats the counterions explicitly that helps include ion-pairing and concominant counterion release in
an efficient way.  The major conclusion we have, that the free ion entropy and bound ion-pair energy dominate the thermodynamic equilibrium, has been shown explicitly and quantitatively by this simple theory. Further, our work includes estimation of the major theormodynamic drives and their quantitative match with simulations for this process of two-chain complexation. 

We believe this model may serve as a basis for building more involved models of complexation and coacervation which will address the additional effects. Further, this basic understanding of the complexation of two homopolymers would certainly shed light on the mechanism of complexation among other pairs of charged macroions such as polylectrolyte-colloid, polylectrolyte-protein, protein-protein etc. Worthy of note here is that the nature of the slope of the free energy with overlap, which determines the feasibilty of the process to move forward unhindered, may be utilized to determine the temporal behavior of complexation from kinetic models.

\begin{section}{acknowledgement}

The authors acknowledge financial support from IISER Kolkata, Ministry of Education, Government of India, and thank Souradeep Ghosh for intense discussions, verification of results, and critical reading of the manuscript.

\end{section}





\bibliography{achemso-demo-main.bib}
\bibliographystyle{unsrt}

\end{document}